\documentclass[12pt,a4paper,notitlepage]{article}
\pdfoutput=1
\usepackage[left=2.5cm, right=2.5cm, top=2.5cm, bottom=2.5cm]{geometry}
\usepackage[section]{placeins}
\usepackage[T1]{fontenc}
\usepackage[utf8]{inputenc}
\usepackage{float}
\usepackage[caption = false]{subfig}
\usepackage{amsmath}
\usepackage{graphicx}
\usepackage{natbib}
\usepackage{setspace,tabularx}
\begin{document}

\author{Tom Broekel\thanks{Tom Broekel, Department of Human Geography and Spatial Planning, Faculty of Geosciences, Utrecht University, t.broekel@uu.nl}
	\\ Utrecht University}
\title{Measuring technological complexity - Current approaches and a new measure of structural complexity}

\date{\today{}}
\maketitle

\begin{abstract}
The paper reviews two prominent approaches for the measurement of technological complexity: the method of reflection and the assessment of technologies’ combinatorial difficulty. It discusses their central underlying assumptions and identifies potential problems related to these. A new measure of structural complexity is introduced as an alternative. The paper also puts forward four stylized facts of technological complexity that serve as benchmarks in an empirical evaluation of five complexity measures (increasing development over time, larger R\&D efforts, more collaborative R\&D, spatial concentration). The evaluation utilizes European patent data for the years 1980 to 2013 and finds the new measure of structural complexity to mirror the four stylized facts as good as or better than traditional measures.
	
\end{abstract}

\thispagestyle{empty}

\newpage
\section{Introduction}



The complexity of technologies is seen as crucial explanatory dimension of technological development and economic success \cite[]{L153, Dalmazzo02}. \cite{Hidalgo2009} argue that country's economic development is shaped by its ability to successfully engage in complex economic activities and technologies. Both \cite{Sorenson2005} and \cite{Balland2015} show that few cities are capable of mastering complex technologies that lay the foundation for their future growth.

Despite its theoretical relevance and an increasing empirical interest, measuring the complexity of technologies empirically is a complicated issue, as \cite{Pintea2007} note: ``\textit{We do not have any easy way to measure complexity}'' [p. 280]. The two most prominent approaches are put forward by \cite{FlSo2001} and \cite{Balland2015}, with the latter transferring the approach of \cite{Hidalgo2009} for approximating economic complexity to the measurement of technological complexity.\footnote{Further approaches can be found in \cite{Albeaik2017} and \cite{FernandezDonoso2017}.} The present paper presents both approaches and argues that they build on the assumption of complexity being scarce at their core. \cite{Balland2015} assume technological complexity to be spatially scare, while \cite{FlSo2001} build on the idea of complex knowledge combinations appearing less frequently than simple ones. It is shown that these assumptions are either theoretically problematic or may induce challenges in the measures' empirical application.

The paper develops an alternative measure of technological complexity, \textit{structural complexity}, which does not relate scarcity and complexity. The paper proceeds by empirically evaluating the approaches (and including two variants of the traditional approaches) against four stylized facts of technological complexity (increasing average complexity over time, more collaborative R\&D, spatial concentration, and larger R\&D efforts). The empirical assessment is made using patent data for Europe between 1990 to 2015. The new measure of \textit{structural complexity} is shown to match the stylized facts similarly or even better than the traditional measures. Similar to the measure of \cite{FlSo2001}, it is not dependent upon the definition of spatial units.

The paper is structured as follows. The next section discusses the traditional approaches of measuring technological complexity. It also introduces the new measure of structural complexity. Section \ref{facts} presents four stylized facts of technological complexity that will serve as benchmarks for the empirical comparison of the traditional and new complexity measures. The set up of the empirical evaluation is subject to Section \ref{evaluation}, the results of which are presented and discussed in Section \ref{results}. Section \ref{conclusion} summarizes the findings and concludes the paper. 

\section{Two traditional and one new measures of technological complexity}\label{theory}
\subsection{(Re-)combinatorial rareness and complexity}

\cite{FlSo2001} approach technological complexity by conceptualizing technological advancement as a search process for knowledge combination.\footnote{This also includes combination).} They assume that the difficulty of combining knowledge represents technological complexity, with more difficult combinations being required to advance more complex technologies. Their second assumption relates past knowledge re-combination frequencies to the current difficulty of combinatorial innovation. On this basis, they construct a measure of technological complexity resembling the (in-)frequency of past knowledge combination such that small frequencies, after controlling for their chances of random occurrence in an N/K framework of  \cite{Kauffmann1993}, translate into low complexity values. In a follow-up study employing US patent data, they substantiate their results by showing that their measure of technological complexity fits well with inventors' perceived difficulty of the inventive combination process \cite[]{Fleming2004}.

However, does the past (in-)frequency of combination really give a clear approximation of the inventive difficulty and thereby of technological complexity? Less frequent combinations may indeed be caused by the difficulty of the according invention process. Yet, it also seems plausible that there is, or has been, little technological or economic interest in such a combination. For instance, it should be relatively easy to integrate the electronic navigation technology used in cars into horse chariots. However, this combination has rarely been realized, if at all, most likely because there is little market potential for it. 

\subsection{The method of reflection approach}

\cite{Balland2015} propose an alternative measure of economic complexity building on the work of \cite{Hidalgo2009}. They transfer the so-called \textit{method of reflection} used by \cite{Hidalgo2009} to assess economic complexity to empirically derive a measure of technological complexity. The method of reflection is based on diversity and ubiquity and assumes that technological complexity is spatially scarce. Diversity is the number of distinct technologies in a region and ubiquity the number of regions specialized in a technology. The proposed index of technological complexity yields high values for technology \textbf{A}, when places specialized in \textbf{A} are also specialized in other technologies that few other places are specialized in. Put differently, a technology will be evaluated as being complex when it belongs to a group of technologies few places specialize in and these specializations appear in the same places. \cite{Balland2015} apply this approach to patent data and estimate the complexity of technologies considering the technological specialization of US metropolitan statistical areas. The authors find that regions commonly associated with technological and economic success (e.g., San Jose, Austin, Bay area) are highly specialized in complex technologies.

There are many arguments supporting the idea of complexity being spatially scarce (see also subsection \ref{facts}). \cite{Hidalgo2009} argue that in order to be successful in complex activities (e.g. in the development of complex technologies), it requires ``nontradable'' spatial ``capabilities'' including ``property rights, regulation, infrastructure, specific labor skills'' \cite[p. 10570][]{Hidalgo2009}. Similarly, concepts like ``learning regions'', ``innovative milieu'', and ``regional innovation systems'' argue that few regions possess location-specific capabilities yielding advantages for technological development \citep{L19, L135, L527, L87, L89, L316}. The findings of \cite{Sorenson2005} add some empirical support to this by showing that 10 to 15 \% of industrial agglomeration can be explained by technological complexity.

However, technologies' spatial distribution may have multiple sources among which complexity is just one. For instance, corporate R\&D facilities are known to be located close to public universities \cite[]{L184}, whose location is largely determined by policy and historical circumstance. The distribution is also impacted by technologies' geographic diffusion, which depends among others on its degree of maturity, popularity, natural conditions, geographic distances, place of origin, and crucially, economic potential \cite{hagerstrand1967, teece1977, Roge95, L253}. Hence, all these factors that are not related to technological complexity may impact technologies' spatial distribution and potentially distort the complexity measure.

Two more issues are related to the assumption of spatial scarcity. First, it makes the measure highly endogenous when analyzing spatial phenomena. For instance, endogeneity is likely to arise when the spatial distribution of technologies is explained with their levels of complexity using a complexity measure based on their spatial distribution \cite[see, e.g.,][]{Balland2015}. Crucially, this issue prevents a sound empirical test of the measure's underlying assumption of complexity being spatially scarce.

Second, as the measure requires a spatial delineation of regions, it becomes conditional on this definition. Put differently, a technology's complexity may depend on the employed spatial unit, i.e., the size of the regions.



\subsection{A  measure of structural complexity}\label{structural_complex}

\cite{FlSo2001} base their measure on ideas of complex systems. I follow this line of thinking and start with technological advancement being a knowledge combination process. I also follow their argument of technologies' complexity being related to the difficulty of combining knowledge pieces in its advancement. Knowledge can be thought of as a ``network'' of knowledge combination with the nodes being individual knowledge pieces and their combination representing the links. To borrow the example of \cite{Hidalgo2015}, think of an \textit{airplane} as a specific type of technology. In order to fly, the airplane combines many different knowledge pieces. Crucially, some pieces need to be directly linked in order to function (e.g., wing design and aluminum processing), while others just need to be indirectly related (e.g., electronic navigation and wing design). When representing the airplane as the network of combined knowledge pieces, wing design and aluminum processing are directly linked. In contrast, electronic navigation is only indirectly related, as other knowledge pieces (electronic control systems, mechatronical interfaces, etc.) act as bridges.

In this conception, I propose to use the complexity of this network representing the combinatorial structure of knowledge pieces as a measure of the (airplane) technology's complexity. That is, the difficulty of combining knowledge is argued to be determined by the precise structure with which knowledge pieces are integrated with each other in innovation processes. Complex structures are more difficult to realize and hence represent more complex technologies. This is motivated by two arguments, one being inspired by the literature on network complexity \citep{Simon1962, bonchev2005} in combination with the literature on knowledge relatedness \cite[]{Noot2000, L702} and the second by information theory \cite[]{Wiener1947, Shannon1948}.

\begin{figure}
	\subfloat[Star]{\includegraphics[width = 2in]{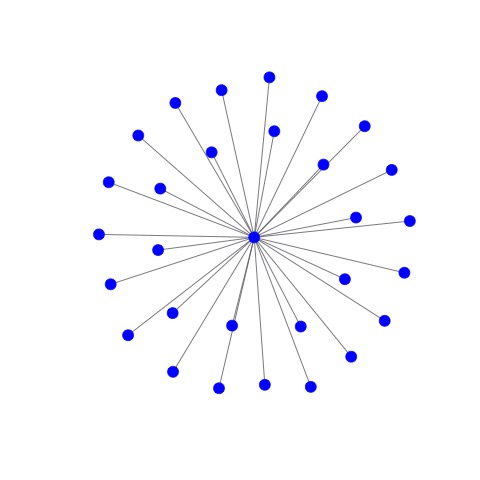}}
	\subfloat[Tree]{\includegraphics[width = 2in]{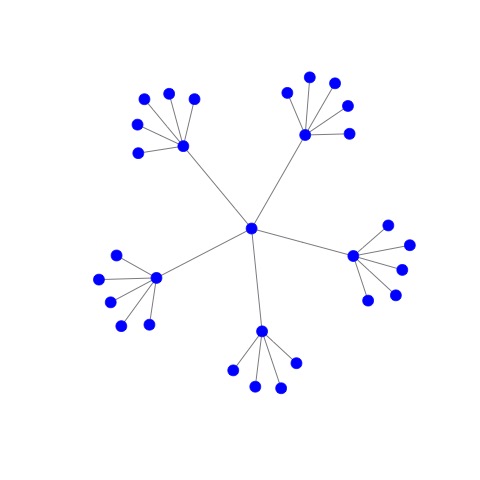}}  
	\subfloat[Small-World]{\includegraphics[width = 2in]{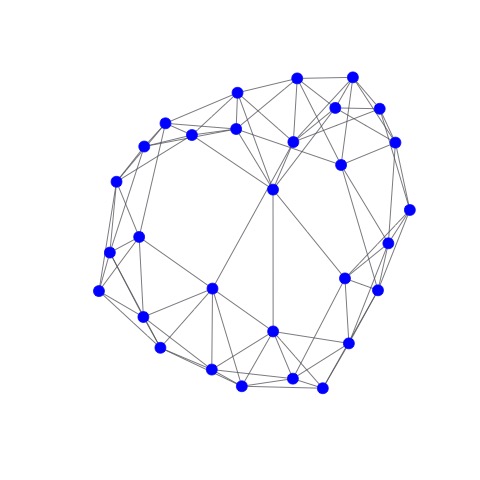}}
	\\
	\subfloat[Complete]{\includegraphics[width = 2in]{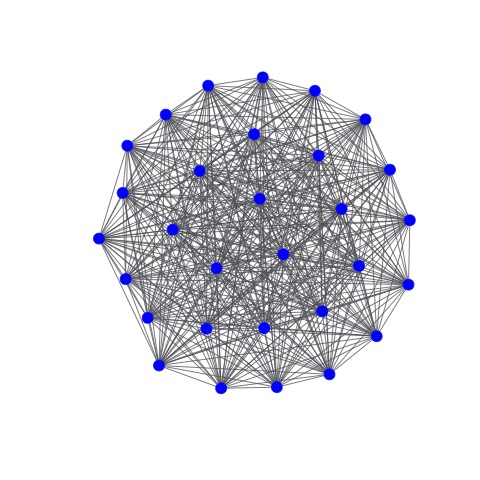}}
	\subfloat[Random]{\includegraphics[width = 2in]{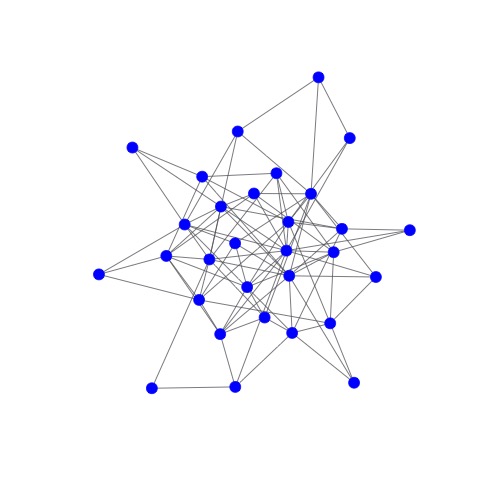}}
	\caption{Typical network structures} 
	\label{examples}
\end{figure}

Beginning with the literature on network complexity and knowledge relatedness, Figure \ref{examples} shows ideal typical network structures. If the combinatorial network has the shape of a star, it means that all knowledge pieces just need to be combined with a central one. As knowledge piece combinations require some technological / cognitive overlap \citep{Noot2000, L69}, the pieces share some common parts making combination/integration easier. The same reasoning applies to fully connected networks (Complete). Such overlap is lower when multiple central knowledge pieces characterize the combinatorial network (tree structure). In this sense, the network resembles the idea of the knowledge space \cite[see, e.g.,][]{Kogler2013, Balland2015}. The greater knowledge diversity makes such network structures more complex. A tree network implies a modular structure with each module being made of somewhat similar knowledge pieces, which reduces the overall complexity in the network. Such clear-cut modules as in a tree network are less frequent than small-world network structures, which therefore indicate greater knowledge diversity. However, there is still a certain degree of modularity and symmetry, which provides some simplifying patterns. Any of those are lacking in purely random combinatorial structures (Random). Each element is combined in a distinct way and there are no overarching principles structuring the combinatorial processes. Complexity is highest in this case. Hence, such structural differences (stars, complete, trees, small-world, random) in combinatorial networks can be used to differentiate complex and simple technologies.

An alternative motivation for using combinatorial networks as way to approach technological complexity is provided by information theory \cite[]{Wiener1947, Shannon1948}. The combinatorial network represents a system of (knowledge) pieces and their interaction (combination). Systems' complexity increases with the amount of information contained in its structure \citep{Dehmer2009}. For instance, a star is simple because it can be summarized by the number of pieces (nodes) and the identity of the central piece (node). Much more information are contained in tree and small-world networks. However, the existence of structuring principles allows for information to be condensed. This is not possible in the case of random (network) structures, which therefore contain maximum information. A complete network is also a simple structure as it represents little information besides the number of pieces \cite[see for a discussion, e.g.][]{bonchev2005,Dehmer2011}. Hence, the information theoretical perspective on  networks also allows for differentiating complex and simple structures and can therefore be used to assess the complexity of combinatorial networks and thereby that of technologies.


Unfortunately, there is no single widely accepted method of measuring the complexity of (combinatorial) network structures. In contrast, a wide range of approaches exists that capture different structural aspects. It is beyond the scope of the present paper to review or discuss their pros and cons \cite[for excellent reviews, see][]{bonchev2005, Dehmer2011, Emmert-Streib2012}.

Recently, \cite{Emmert-Streib2012} developed the so-called \textit{Network Density Score} (NDS), which reflects the structural diversity in a network. The measure has a number of desirable features. Most importantly, it convincingly differentiates ordered, complex, and random networks. Networks are considered \textit{ordered} when many nodes show similar properties (e.g., degree). For instance, most nodes in a star and tree network have the same degree (one). According to the above discussion, ordered networks represent simple technologies because they contain less information and are more homogeneous. Complex networks represent mixtures of such ordered and random structures while random networks lack any type of order. In accordance with the above, complex networks belong to less complex technologies than random networks. \cite{Emmert-Streib2012} show that no traditional measure of network complexity is similarly good at categorizing networks with respect to their structural complexity. In addition, the $NDS$ measure is relatively invariant to the size of networks; a rather unique feature among the measures of network complexity. It will be shown later in this paper that the measure's size invariance is a strong asset. 

\section{Stylized facts about technological complexity}\label{facts}

Each of the approaches of measuring technological complexity takes a somewhat different perspective, so the following question arises: which reflects technological complexity most appropriately? Unfortunately, there is no objective standard against which such a comparison can be made. I therefore put forward a (non-exclusive) list of four stylized facts about technological complexity, which most scholars in the field seem to agree upon. The three approaches will be evaluated on how well the complexity measures constructed on their basis are able to empirically reflect these facts.

\paragraph{Technological complexity increases over time.} Technological systems have become increasingly complex over time because of knowledge and technologies' cumulative nature, with each generation building upon the technological environment established by its predecessors \cite[]{NelsWint82,Howitt1999,Aunger2010, Hidalgo2015}. Technologies also become more complex due to their growing range of functions. For instance, ``[d]\textit{igital control systems }[of aircraft engines] \textit{interact with and govern a larger (and increasing) number of engine components than} [previous]\textit{ hydromechanical ones} \cite[]{Prencipe2000}. Another example is Microsoft's operation system Windows, that grew from 3-4 million lines of code (Windows 3.1) to more than 40 million (Windows Vista) \cite[]{wiki_lines_code_2017}. Moreover, technologies have reached higher levels of complementary requiring more multi-technology activities, which adds to the complexity of their development and application \cite[]{Fai2001}. ``\textit{The result is a constantly increasing sophistication and richness of the technological world}'' \cite[p. 773]{Aunger2010}. The pattern of increasing technological complexity over time should hence be reflected by complexity measures applied to empirical data.

\paragraph{Complex  technologies require more R\&D}

The development of complex technologies requires dealing with greater technological diversity and combining less common knowledge than simple technologies \cite[]{FlSo2001}. Creating new knowledge combinations implies search activities for potentially fitting pieces and subsequent testing of these combinations. Frequently, advancing complex technologies is achieved by trial-and-error \cite[]{Carbonell2006}. ``\textit{What succeeded and failed last time gives clues as to what to try next, etc.} \cite[p. 464]{L164}. Hence, ``harder-to-find'', i.e., more difficult/complex, solutions involve more \textit{trials} and \textit{errors}, which consume resources. The greater knowledge diversity inherent to complex technologies further demands more diverse but specialized experts working together. ``\textit{When dealing with technological complex projects }[...]\textit{, they} [...] \textit{depend more heavily on other functional specialists for the expertise}'' \cite[p. 226]{Carbonell2006}. They must have to be provide with a environment that puts them into position to exchange knowledge, learn, and work together, which requires further (e.g., organizational) resources \cite[]{Teec92}. In particular, (spatial) proximity among experts allowing for face-to-face communication enhances the work on complex projects, which is not necessarily true for simple projects in which intensive communication may even have negative effects \cite[]{Carbonell2006}. Related to these are the greater difficulties of transmitting and diffusing more complex knowledge \cite[]{SoRF2006a}. Learning of complex knowledge is more resource-intensive because greater absorptive capacities are needed \cite[]{L104} and passive learning modes are insufficient \cite[]{Pintea2007}. This challenges communication and collective learning processes within and among R\&D labs.

While there is no direct empirical confirmation for this stylized fact, some findings support it. For instance, the development time of complex products is larger (and hence more expensive) than that of simple ones \cite[]{Griffin1997}. Studies also find nations' R\&D intensities outgrowing their economic outputs and incomes \cite[]{Pintea2007,WooKim2015}. The  greater need of collaborative R\&D in case of complex technologies is also frequently related to larger resource requirements that are overcome by organizations pooling their resources \cite[see, e.g.,][]{Hage00}. Moreover, the larger uncertainty and costs associated to complex technologies makes organizations engaging in their development more likely to fail \cite{singh1997}.

\paragraph{Complex technologies require more cooperation.} ``\emph{With the universe of knowledge ever expanding, researchers need to specialise to continue contributing to state of the art knowledge production}'' \cite[p. 723]{Hoekman_et_al_2009}. This in turn has led to a stronger dispersion of knowledge in the economy, thereby increasing the relevance of interpersonal knowledge exchange. Put differently, technological advancement increasingly requires interpersonal interaction and cooperation \cite[]{Meyer2004, WuJU2007}. This trend is reflected in empirical data. For instance, \cite{WagnerDoebler2001} show that about ten percent of scientific publications were realized by co-authorships at the beginning of the twentieth century. This percentage rose to almost fifty percent at the end of this century. A similar trend can be observed for patents \cite[]{FlFr2007}. Interaction and cooperation is thereby more crucial for the development of complex than simple knowledge, as complex technologies include the combination of diverse and heterogeneous knowledge \cite[]{L253}. These are more likely possessed by specialized experts \cite[]{Hidalgo2009, Hidalgo2015, Balland2015}. This finds some indirect confirmation in the studies of \cite{KatzMartin1997} and \cite{Frenkenetal2005}. These authors report positive correlations between the number of citations to scientific articles (as a rough measure of their quality) and their numbers of authors.

\paragraph{Complex technologies concentrate in space.} As has been argued for a long time in Economic Geography and Regional Science as well as more recently by \cite{Hidalgo2009} and \cite{Balland2015}, developing complex technologies requires special skills, existing expertise, infrastructure, and institutions not found in every place. For instance, industrial sectors interlinked by labor mobility, open but dense social networks, and related knowledge bases are crucial factors in such contexts \cite[]{Saxe94, Castaldi2015}.  Adding to this are strong economies of scale in R\&D and the location choice of large R\&D labs and universities that tend to be highly agglomerated \cite[]{Jaff1989a, audretsch_feldman_1996, Almeida1996}. The place-specificity of favorable conditions for innovation are emphasized in concepts like the ``learning regions'', ``innovative milieu'', and ``regional innovation systems'' \citep{L87, L89, L316}. These conditions allow for bridging cognitive distances and combining heterogeneous knowledge, which in other places would remain uncombined. Such conditions are path-dependent and place-specific making places with such characteristics relatively rare. The studies of \cite{Balland2015} and \cite{Sorenson2005} confirm this stylized fact using U.S. patent data.



\section{Empirical evaluation}\label{evaluation}

To compare the approaches of measuring technological complexity, I will estimate five measures and apply them to empirical data. Subsequently, I will evaluate if the obtained results meet the four stylized facts above.

\subsection{Data}\label{data}

In a common manner, I rely on patent data for approximating knowledge and technologies. Despite well-known problems \cite[see for a discussion][]{L589}, patents entail detailed and unparalleled information about innovation processes such as date, location, and a technological classification. I use the OECD REGPAT database covering patent applications and their citations from the European Patent Office. The data covers the period 1975 to 2013 and  includes information on $2.823.975$ patent applications. I remove all non-European inventors leaving $1.393.411$ patents that are assigned to European NUTS 2 and 3 regions by means of inventors' residence (multiple-counting).

Technologies are defined on the basis of the \emph{International Patent Classification} (IPC). The IPC is hierarchically organized in eight classes at the highest and more than 71,000 classes at the lowest level. I use the four-digit IPC level to define 630 distinct technologies. While there is no objective reason for this level, it offers a good trade-off between technological disaggregation and manageable numbers of technologies. In addition, it has been used in related studies \cite[]{L392,L762}.

The complexity measures are estimated in a moving window approach. Patent numbers vary considerably between years and some technologies have few patents. I therefore follow common practice and combine patent information of five years such that a complexity measure estimated for year $t$ is based on patents issued between $t$ and $t-4$ \cite[see, e.g.,][]{TerWal2013}.

\subsection{Estimation of complexity measures}
\subsubsection{Measures based on the method of reflection}\label{reflection}
The estimation of the complexity measures based on the method of reflection starts with the calculation of the regional technological advantage (RTA) of region $r$ with respect to to technology $c$ in year $t$.
\begin{equation}
RTA_{r,c,t}=\frac{\frac{patents_{r,c,t}}{\sum_r patents_{r,c,t}}}    {\frac{\sum_{c} patents_{r,c,t}}{\sum_{c} \sum_{r} patents_{r,c,t}}}
\end{equation}
Second, an incidence matrix (\textbf{M}), or two-mode network, between regions (rows) and technologies (columns) is constructed with a binary link if region $r$ has $RTA_{r,c,t}>1$, i.e., it is above average specialized in technology $c$, and no link otherwise. Each region's number of links (row sum) represents its diversity ($K_{r,0}$) and each technology's links its ubiquity ($K_{c,0}$) (column sum). In accordance with \cite{Hidalgo2009}, the diversity and ubiquity scores are sequentially calculated by estimating the following two equations simultaneously over $n$ (20) iterations \cite[for more details, see][]{Balland2015}.
\begin{equation}
KCI_{r,n}=\frac{1}{K_{r,0}} \sum_{r} M_{r,c} K_{r,n-1}
\end{equation}
\begin{equation}
KCI_{c,n}=\frac{1}{K_{c,0}} \sum_{c} M_{r,c} K_{c,n-1}
\end{equation}
In the present paper, I am particularly interested in $KCI_{c,n}$, which represents technologies' complexity value. As a robustness check, the complexity index is estimated using the assignments of patents to NUTS 3 (1.383) regions, denoted as $HH.3NUTS$, and alternatively to NUTS 2 (384) regions, which will be denoted as $HH.2NUTS$.

On the basis of the work of \cite{Caldarelli2012} and \cite{Tacchella2012}, \cite{Balland2015} propose an alternative version of this complexity measure. Matrix $M$ is column standardized and multiplied with its transposed version to get the square matrix $B$, which has the 630 technologies as dimensions. Its none-diagonal elements represent the similarity of technologies' distributions across places. The diagonal is the average diversity of cities having an RTA in the row/column technology. A technological complexity score is then estimated as the second eigenvector of matrix $B$. It is called $HH.eigen$.

Accordingly, two measures are based on the original method of reflection ($HH.3NUTS$, $HH.2NUTS$) that vary in terms of the underlying spatial unit. In addition, a modified version of the method of reflection is used for the measure $HH.Eigen$.\footnote{The three measures have been estimated using the R-package EconGeo by \cite{Balland2017}.}

\subsubsection{Measures based on the difficulty of knowledge combination}

For calculating the complexity measure of \cite{FlSo2001}, knowledge pieces need to be defined whose combinations can then be evaluated. In accordance with \cite{FlSo2001}, knowledge pieces are approximated by the most disaggregated level of IPC subclasses (ten-digit subclass IPC level). Knowledge combinations are these subclasses' co-occurrences on patents (patents are usually classified into multiple classes). The ease of combination is approximated by setting the co-occurrence count of subclass $i$ with all other subclasses in relation to the number of patents in this subclass. 
\begin{equation}
E_i=\frac{count\; of\; subclasses\; previously \;combined \;with\; subclass \; i}{count\; of\; previous\; patents \;in \; subclass \;i}
\end{equation}
This score is inverted and averaged over all patents of subclass $i$ to create a measure of independence for each patent.
\begin{equation}
K_l=\frac{count\;of\; subclasses \;on\; patent\; l}{\sum{E_i}_{l \epsilon i}}
\end{equation}
Based on the N/K model of \cite{Kauffmann1993}, the final complexity score is estimated as the ratio between the measure of independence $K_l$ and the total number of patents on which $l$'s occurs ($N$). Crucially, $E_i$ and the count of subclasses on patent $l$ are estimated on the basis of different time periods. While the latter is calculated with respect to the current time period (moving window: patents granted between $t$ and $t-4$), the first considers all patents prior to $t-4$. The score is estimated for each patent and subsequently averaged across all patents belonging to a technology (four digit IPC class). It is denoted as $FS.Modular$.

\subsubsection{Calculation of the measure of structural complexity}
The calculation of the new measure of structural complexity ($Structural$) begins in a similar manner as $FS.Modular$. First, for each of the 630 technologies $c$, the set of patents are extracted belonging to the respective class. Second, the matrix $M_c$ is established for each set by counting all co-occurrences of (ten-digit) IPC subclasses on its patents. $M_c$ is dichotomized with all positive entries being set to one. The matrix now represents a binary undirected network $G_c$ with the nodes being all IPC subclasses occurring on patents with at least one IPC subclass belonging to technology $c$. Links indicate observed co-occurrence. $G_c$ contains all ways technology $c$'s subclasses have been combined among themselves and with all other patent subclasses. Hence, it is the combinatorial network of technology $c$.\footnote{Alternatively, the network can be restricted to subclasses belonging to technology $c$. However, such approach would ignore potential bridging functions of adjacent technologies as well as the possibilities of embedding this technology into larger technological systems.} The question now is whether this network $G_c$ has a \textit{complex} structure.

The network complexity $NDS$ measure of \cite{Emmert-Streib2012} provides an answer. In contrast to most traditional network complexity measures, the NDS combines multiple network variables into one. First, the share of modules in the network ($\alpha_{module}=\frac{M}{n}$) with $M$ being the number of modules and $n$ that of nodes. Modules can be seen as sign of general organizational principles in the network, i.e. of the existence of ordered structures. Second, a measure of the variance of module sizes $v_{module}=\frac{var(m)}{mean(m)}$, whereby $m$ is the vector of module sizes. It approximates ``\textit{the variability of network sizes in respect to the mean size of a module}'' \cite[p. e34523]{Emmert-Streib2012}. Random networks are likely to show a low variability and low average size of modules. Third, the variable $V_{\lambda}$ capturing the Laplacian ($L$) matrix's variability is defined as $ v_{\lambda}=\frac{var(\Lambda(L))}{mean(\Lambda(L))}$, which picks up similar structures as $v_{module}$. Fourth, the relation of motifs of size three and four ($r_{motif}=\frac{(N_{motif} (3))}{N_{motif} (4)}$). In numerical exercises \cite{Emmert-Streib2012} observe this variable to be highest in ordered, medium in complex, and lowest in random networks. 

The four variables are combined in order to obtain the individual network diversity score ($INDS$) for the network ($G_c$):
\begin{equation}\label{inds}
INDS(G_c)=\frac{\alpha_{module}*v_{module}}{v_{\lambda}*r_{motif}}.
\end{equation}

Networks may show properties of a complex or ordered network just by chance and thereby mislead measures of complexity. \cite{Emmert-Streib2012} therefore estimate $INDS$ for a population of networks $G_M$, to which $G_i$ belongs. In practice, this is achieved by drawing samples $S$ from network $G_c$ and estimating $INDS$ for each sample network. The final network diversity measure ($NDS_s$) can than be obtained by:
\begin{equation}\label{nds}
NDS_s (\{ G_c^S | G_M \}) = \frac{1}{S} \sum_{G_c \epsilon G_M}^{S} INDS(G_c)
\end{equation}

Since the network density score (NDS) is only defined for sufficiently large and connected networks \cite[]{Emmert-Streib2012}, I restrict the estimation to the largest component of network $G_c$. Moreover, the $NDS_c$ score (equation \ref{nds}) is only calculated if the component has at least five nodes (co-occurring IPC subclasses). More precisely, for each $G_c$ (main component), a sample of 100 nodes $n$ (in case of components with less than 1.000 nodes) and 300 (for components with more than 1.000 nodes) is randomly drawn. For each node $n$, a network $G_n$ is drawn from $G_c$ by a random walktrap of 1.000 steps starting from $n$. From this network, a subnetwork $G_n^i$ of 200 random nodes $i$ \footnote{\cite{Emmert-Streib2012} find a sample network size of 120 nodes to be sufficient for robust results.} is selected. $INDS$ (equation \ref{inds}) is then estimated for $G_n^i$. The score is subsequently averaged over all subnetworks giving $NDS_c$. To obtain values with large values signaling random networks (complex technologies), medium values indicating complex networks (medium complex technologies), and low values standing for ordered networks (simple technologies), $NDS_c$ is taken in logs and multiplied by $-1$. It represents the structural (combinatorial) complexity of technology $c$ and is denoted as $Structural$. Notably, the results (i.e., the ranking of technologies) will somewhat vary by default when the measure is repeatedly estimated \footnote{The estimations of the measures' parts have been conducted with the R-package QuACN by \cite{Mueller2011}} due to the measures' random component.

\section{Results} \label{results}
\subsection{Application oriented aspects of the complexity measures}

Before the measures are evaluated against the four stylized facts, it is informative to examine some empirical features unrelated to the four stylized facts. Unfortunately, two technologies do not have sufficient patents for any measures to be estimated leaving sample of 628 technologies in the example year 2010. Sixteen lack a sufficiently large component in the combinatorial network for a calculation of structural complexity. Table \ref{descriptives} in the Appendix lists some basic descriptives.

A first interesting insight into the measures' properties is gained by rank-correlation analyses using the data of the last five years (2008-2013) (Table \ref{correlation_recent}). Besides the five complexity measures, the analyses include the growth of patents in the last 10 years (Patent.Growth.10), the number of citations per patent (Cit.Pat), and the number of IPC subclasses (IPCs) found on patents of a technology.	

\begin{table}[ht]																				
	\centering
	\resizebox{\textwidth}{!}{%
		\begin{tabular}{rccccccccc}							
\hline																					
		&Patents & Patents	&	Cit.Pat		&	IPCs	&	HH.	&	HH.	&	HH.	&	FS.	& Structural\\
		&			 & 	Growth.10	&		&	&	NUTS3	&	NUTS2	&	Eigen	&	Modular	& \\
	\hline
Patents & 1.00 & 0.12 & 0.01 & 0.78 & -0.56 & -0.53 & 0.06 & 0.39 & 0.68 \\ 
Patent.Growth.10 & 0.12 & 1.00 & -0.02 & -0.00 & -0.03 & -0.06 & 0.11 & 0.08 & 0.07 \\ 
Cit.pat & 0.01 & -0.02 & 1.00 & -0.03 & 0.12 & 0.14 & -0.01 & 0.07 & 0.00 \\ 
IPCs & 0.78 & -0.00 & -0.03 & 1.00 & -0.42 & -0.38 & 0.08 & 0.13 & 0.37 \\ 
HH.3NUTS & -0.56 & -0.03 & 0.12 & -0.42 & 1.00 & 0.89 & 0.43 & -0.14 & -0.47 \\ 
HH.2NUTS & -0.53 & -0.06 & 0.14 & -0.38 & 0.89 & 1.00 & 0.38 & -0.20 & -0.43 \\ 
HH.Eigen & 0.06 & 0.11 & -0.01 & 0.08 & 0.43 & 0.38 & 1.00 & 0.07 & -0.03 \\ 
FS.Modular & 0.39 & 0.08 & 0.07 & 0.13 & -0.14 & -0.20 & 0.07 & 1.00 & 0.15 \\ 
Structural & 0.68 & 0.07 & 0.00 & 0.37 & -0.47 & -0.43 & -0.03 & 0.15 & 1.00 \\ 
	\hline
\end{tabular}
}
\caption{Correlation of complexity measures 2009-2013}
\label{correlation_recent}																				
\end{table}	

No measure shows a strong relationship with the number of citations per patents (Cit.Pat). Research shows a correlation between patents' technological and economic values with their citation counts \cite[see, e.g.,][]{Traj90, HNSV1999} suggesting that no measure seems to be able to directly capture this ``value'' dimension of technologies. Similar holds true for the growth of patent numbers during the last 10 years (Patents.Growth.10).

$HH.2NUTS$ and $HH.3NUTS$ are positively correlated. Their correlation is relatively high with $r=0.89$ implying that the employed scale of the underlying spatial units matters but does not dramatically alter the complexity scores. Therefore, one of the criticisms of this measure raised in Section \ref{reflection} find weak support. Put differently, the ranking of technologies in terms of complexity depends to some but not to a dramatic degree on the spatial unit chosen as the basis in the estimation.  

The two measures based on IPC subclass combinations ($FS.Modular$ and $Structural$) are negatively associated with the other complexity measures (except for $FS.Modular$ and $HH.Eigen$). Accordingly, while attempting to measure the same thing (technological complexity), the two approaches (method of reflection and evaluating IPC subclass combinations) do not overlap empirically.

It should be noted that the computational requirements of $Structural$ drastically exceed those of the other measures. In part, this is due to the fact that it is not yet implemented in existing software and (more significantly) it includes an iterative procedure.

\subsection{Increasing complexity over time}
\subsubsection{Average complexity}

\begin{figure}[ht]
	\centering
	\includegraphics[width=0.8\textwidth]{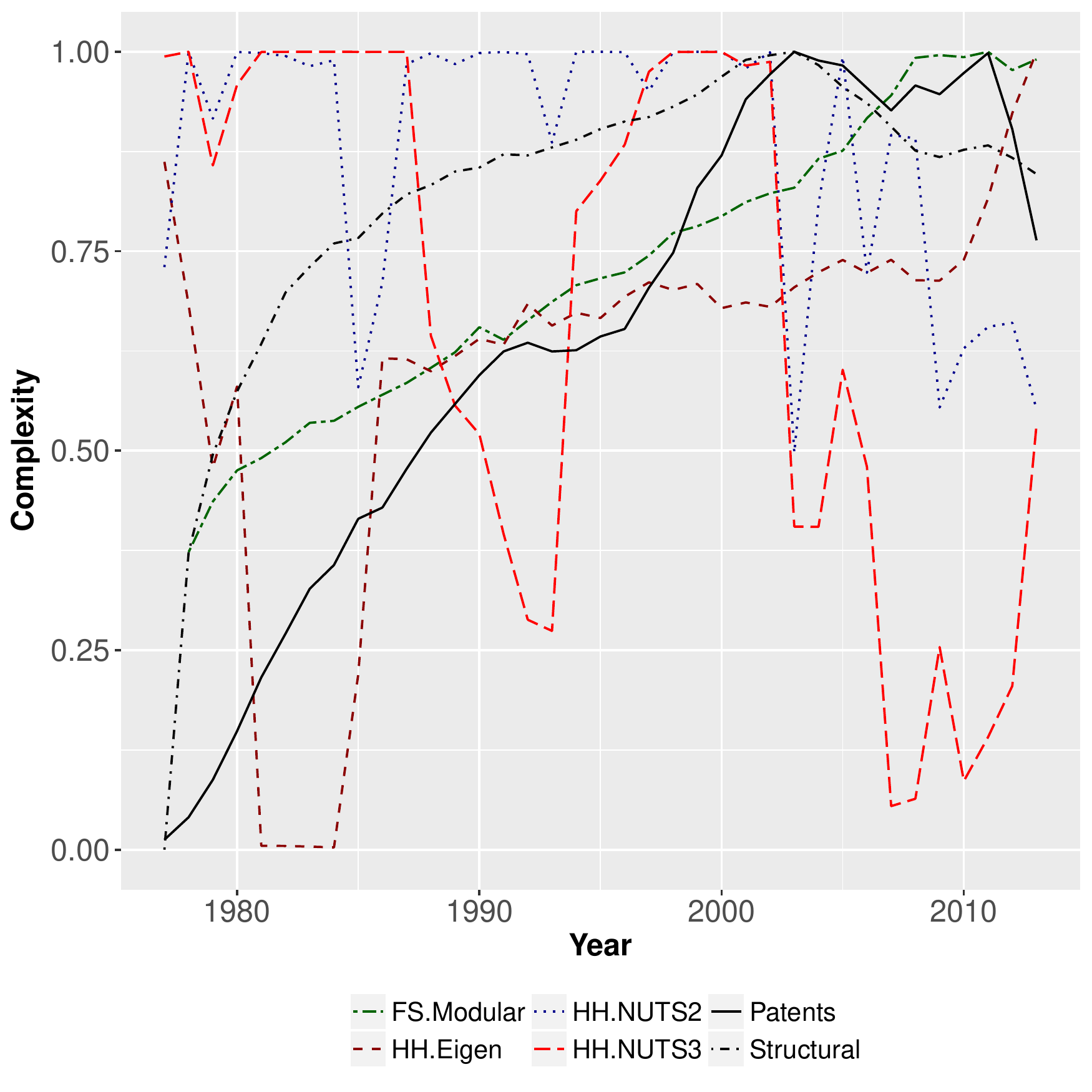}
	\caption{Average complexity 1980-2013}
	\label{over_time}
\end{figure}

While the application-oriented aspects are important, they don't give insights into how well the different approaches perform in measuring technological complexity. The first stylized fact used for such an assessment is whether the average complexity of technologies increases over time. Figure \ref{over_time} answers this question by showing the median complexity value across all technologies for each of the five measures from 1980 to 2013. For better visualization and comparison, all measures have been divided by their maximum.
The first thing to notice is the relatively erratic and nonparallel development of $HH.2NUTS$ and $HH.3NUTS$. With some interruptions, $HH.3NUTS$ remains close to one (maximum) until about 2000, before it starts to drop to values around 0.55. In contrast, $HH.3NUTS$ starts from a maximum value of almost one, before dropping to about 0.27 in 1993, increasing back to one in 1997, and declining again strongly until 2008, before growing in the last three years. While technological development does not necessarily take place in a smooth manner, there are no explanations for why complexity should have dropped that drastically at some point in time. Moreover, the nonparallel development of $HH.2NUTS$ and $HH.3NUTS$ underlines the scale variance of the measure. Clearly, the two measures fail in representing the stylized fact of increasing complexity over time.

The three other measures, $HH.Eigen$, $FS.Modular$, and $Structural$ are more effective. While there is a strong drop in $HH.Eigen$ to almost zero in the early 1980s, it increases relatively monotonically afterwards. $FS.Modular$ and $Structural$ show a more steady and monotonic increase, which however turns in the year 2004 in case of $Structural$. The decline of $Structural$ is rather limited (the value of 2013 is just 7.3 \% smaller than the maximum value in the year 2004). The decline might be a feature of the employed database where recent patents are frequently added multiple years after their actual application and hence they might not have been included yet. It should therefore not be over interpreted. 
In general, the figure shows the similarity in the developments of $FS.Modular$ and that of the median number of patents per technology (also normalized with its maximum). $Structural$ follows the general trend of patent numbers as well but to a lower degree. The extent to which this might be caused by a ``size dependency'' of the complexity measures, will be explored in more detail in Section \ref{magnitude}. 

\subsubsection{Technologies' age}
Increasing complexity over time can also be assessed by comparing the average `age' of technologies to their complexity, with the idea being that more recent technologies are more complex. I approximate age by calculating the mean age of patents in a given year for each technology and correlate it with the according complexity scores.\footnote{Note that the database is restricted with the earliest patents being from 1978.} A positive correlation implies that technologies with young patents (e.g., subject to more recent R\&D) obtain higher complexity values, which corresponds to the stylized fact.

\begin{figure}[ht]
	\centering
	\includegraphics[width=0.8\textwidth]{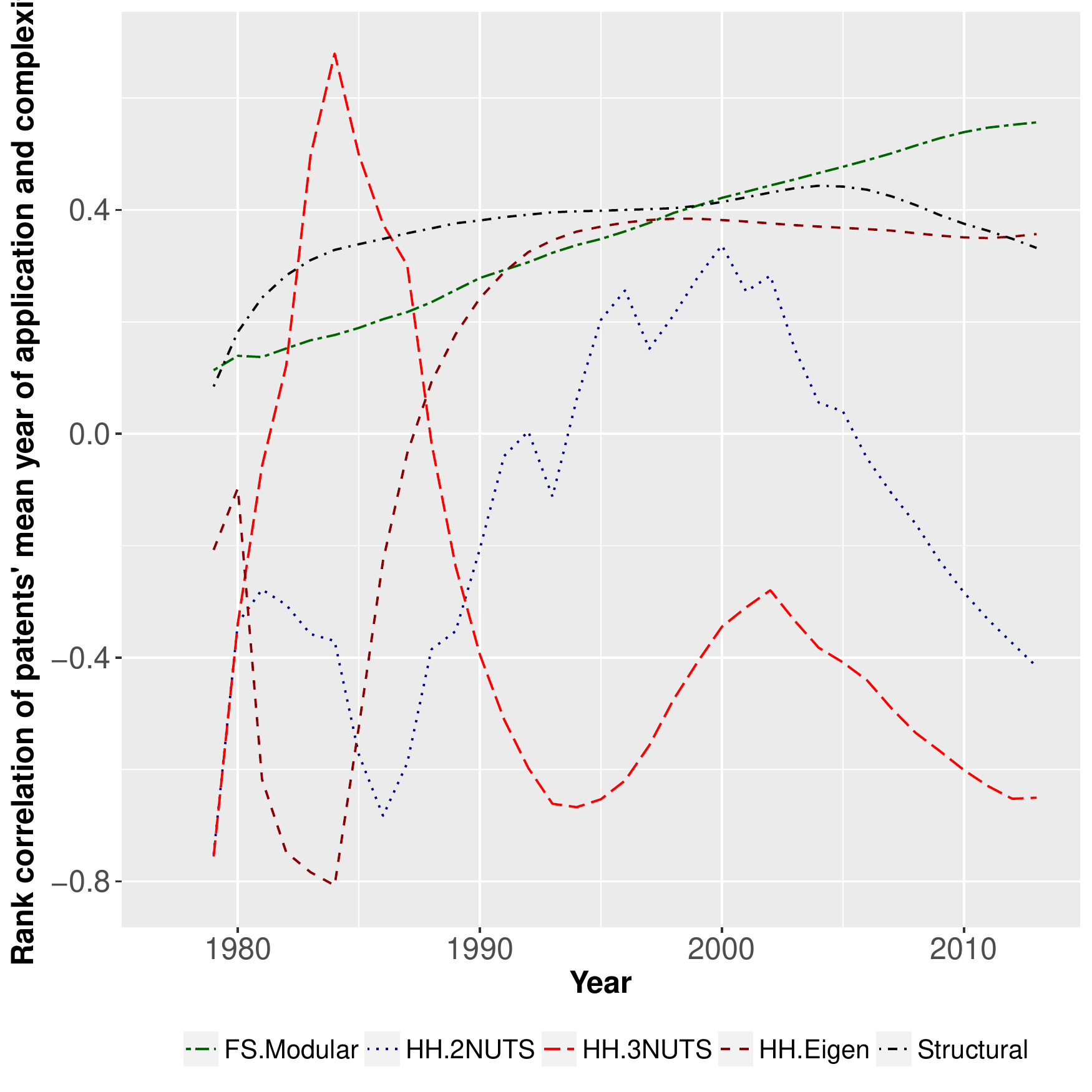}
	\caption{Correlation with patents' mean age, 1980-2013}
	\label{mean_age}
\end{figure}

Figure \ref{mean_age} plots this rank correlation for each year. It clearly confirms the previous observation: just $HH.Eigen$, $FS.Modular$, and $Structural$ are able to replicate the stylized fact of younger technologies being more complex, i.e., growing in complexity over time. Notably, the correlation of $HH.Eigen$ and patents' mean age only becomes positive after 1986, while for $FS.Modular$ and $Structural$ it has been positive since 1981.\footnote{Given the lack of patent data prior to 1978, early years may not be reliable for this analysis.} $HH.2NUTS$ and $HH.3NUTS$ are characterized by a negative correlation for most years suggesting that they identify older technologies as complex.

In summary, the three measures $HH.Eigen$, $FS.Modular$, and $Structural$, correspond to and reflect growing technological complexity over time and thereby align with the first stylized fact.

\subsection{Magnitude of R\&D efforts}\ref{magnitude}

Unfortunately, I lack information on the true R\&D efforts invested or R\&D employment contributing to the development of the technologies considered in the paper. In a common manner, I therefore approximate the R\&D efforts with the number of patents. This is justified by patents and R\&D efforts being positively correlated at the organizational and regional level \cite[]{L589, L539}. However, it has to be pointed out that this approximation is strongly influenced by national and industrial differences in patent propensity and R\&D productivity \cite[for a discussion, see ][]{L265,DeRassenfosse2009}. This surely reduces the reliability of the analysis and calls for future work on this issue.\footnote{Alternatively, I could have used the number of inventors as approximation of R\&D efforts. However, their correlation with patent counts is $r=0.98***$ and does not impact the empirical results at all.} 

\begin{figure}[ht]
	\centering
	\includegraphics[width=0.8\textwidth]{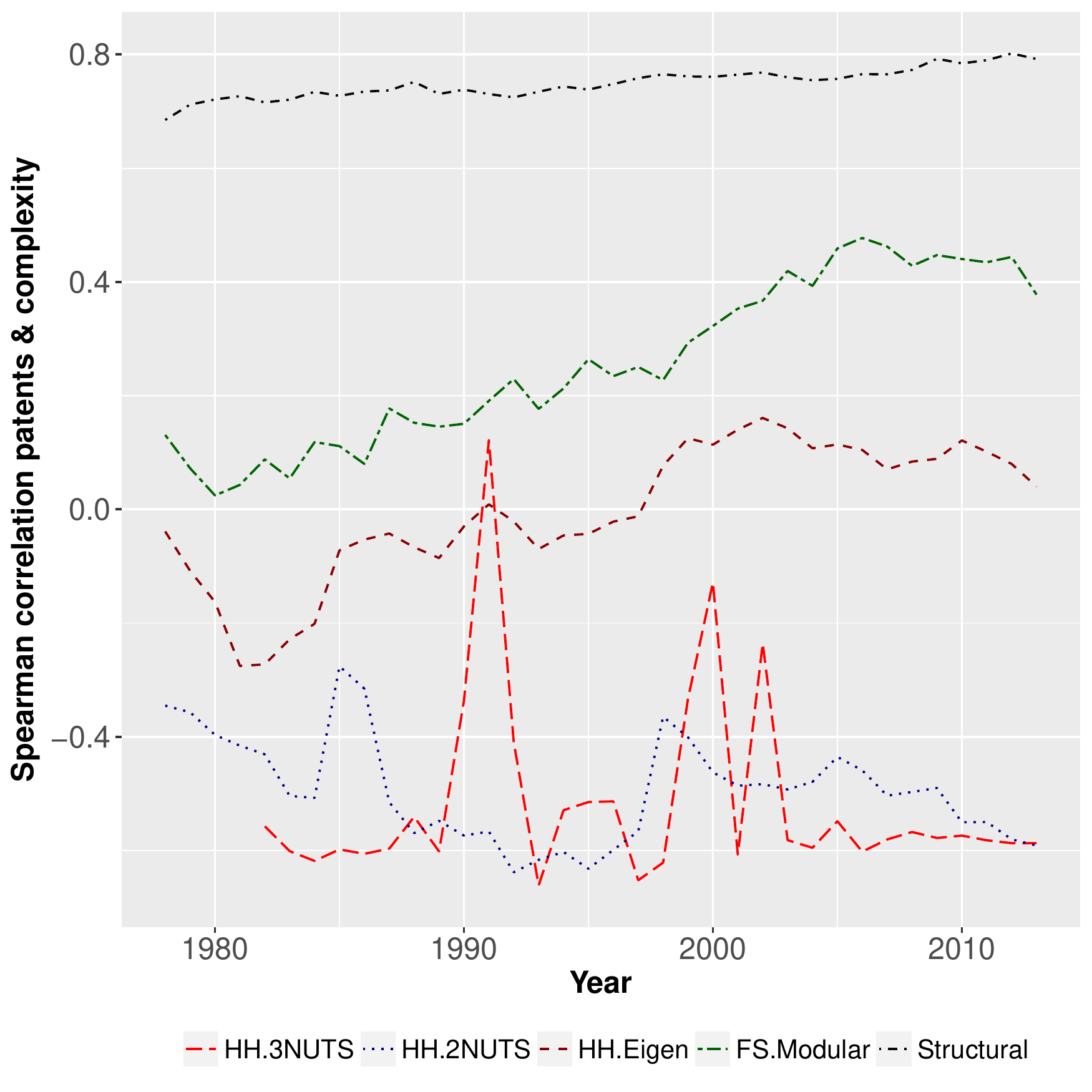}
	\caption{Correlation of complexity with technologies' patent counts, 1980-2013}
	\label{size}
\end{figure}

The results of the (rank) correlation analysis are shown in Figure \ref{size}. The two measures $HH.3NUTS$ and $HH.2NUTS$ are strongly negatively correlated with patent counts for all years, except 1991. The negative correlation of $HH.3NUTS$ and $HH.2NUTS$ may reflect that technologies with few patents tend to be (for this reason) (co-)concentrated in space, which will increases their estimated complexity. The strong negative correlation implies that these two measures cannot resemble this stylized fact.

\begin{figure}[ht]
	\centering
	\includegraphics[width=0.8\textwidth]{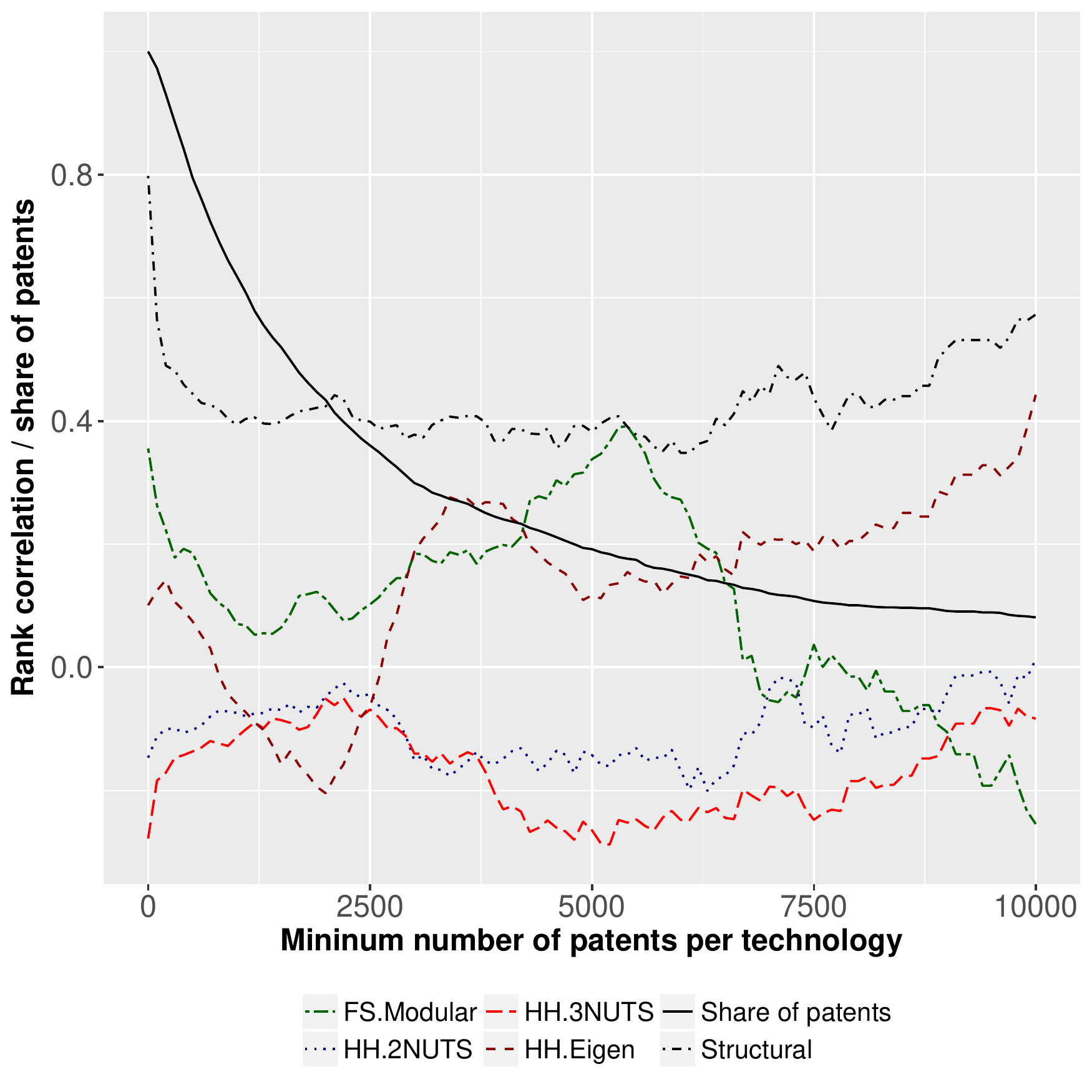}
	\caption{Correlation of complexity with patent counts across size classes}
	\label{size_class}
\end{figure} 

A positive correlation between patent numbers and complexity scores are observed for $Structural$. Large patent classes imply many IPC subclasses ($r=0.93***$), which reduces the chances of their co-occurrence on patents. The correlation of $Structural$ is above $0.6$ in most and above $0.8$ in recent years. Hence, the measure seems to be strongly influenced by the number of patents assigned to 4-digit IPC classes. This makes the measure reflecting this stylized fact easily.n 

However, it also leads to the question whether the measure's information content goes sufficiently beyond that represented by the  absolute number of patents. While the ranking information is not identical, it overlaps to more than 80\%. Figure \ref{size_class} reveals that the magnitude of the correlation drops strongly when very small technologies are excluded. For instance, when excluding patents in technologies with less than 200 patents, which correspond an exclusion of 8\% of all patents, the correlation of $Structural$ and technology size already drops to $0.5$. Given that the measure is based on network complexity measures that are known to be closely linked to networks size, a rank correlation of less than $r=0.5$ has to be seen as a relatively low value in this context and highlights one of the NDS measure's attractive features \cite[]{Emmert-Streib2012}. By further limiting the sample to patents in large technologies, the correlation decreases to a minimum of $0.35$ before gradually increasing again. Crucially, the correlation always remains positive without reaching the initial large levels again. The declining correlation for larger technologies relates to the fact that small technologies with very few patents frequently show complete combinatorial networks (density of 1), which are per definition classified as being simple (see Section \ref{structural_complex}). In sum, the stylized fact can be clearly confirmed for $Structural$. 

A more moderately positive correlation is found for $F.S.Modular$ signaling that this measure clearly represents the stylized fact of complex technologies requiring larger R\&D efforts. Figure \ref{size_class} reveals that this correlation is somewhat larger in case of medium sized technologies than in case of smaller and larger ones.

The results for $HH.Eigen$ are less clear. Its correlation with patent counts remains negative until 1997. Afterwards it becomes positive. Given the positive correlation staying well below $r=0.2$, I argue of this measure aligning to this fact.


In short, only two out of five measures ($FS.Modular$ and $Structural$) are able to mirror the stylized fact of complex technologies being associated to larger R\&D efforts. 

\subsection{Spatial concentration}

The production of complex technologies is expected to be spatially concentrated because few places possess the necessary capabilities. To test this stylized fact, I first estimate the spatial concentration of technologies by means of the GINI coefficient and the assignment of inventors to NUTS3 regions. The coefficient obtains a values close to one if inventors concentrate in few regions and its value converges to zero if they are evenly distributed in space. As a simple test of the degree of spatial concentration, I estimate the correlation between complexity scores and GINI coefficients of the patents used in their construction for the year 2010. The results are shown in Table \ref{spatial correlation}. 

The two measures $HH.3NUTS$ and $HH.2NUTS$ turn out to be strongly positively correlated to spatial concentration, while $HH.Eigen$, $FS.Modular$, and $Structural$ are found to be negatively correlated. 

While this would suggest that just the first two measures correspond to the stylized fact, it has to be pointed out that spatial concentration is strongly negatively correlated with technologies' size (number of patents). Larger technologies concentrate less in space. Since $FS.Modular$ and $Structural$ are positively correlated with size, this is might drive the results.

\begin{table}[ht]																				
	\centering
	\resizebox{\textwidth}{!}{%
		\begin{tabular}{lcccccc}	
			
			& Patents &	HH.3NUTS	&	HH.2NUTS	&	HH.Eigen	&	FS.Modular	&	Structural	\\
			\hline											
			$r$ with GINI coef. & $-0.86^{***}$ & $0.33^{***}$	&	$0.18^{***}$	&	$-0.13^{***}$	&	$-0.37^{***}$	&	$-0.67^{***}$	\\
			\hline											 																			
		\end{tabular}
	}
	\caption{Correlation between inventors' spatial distribution and technological complexity in 2010}
	\label{spatial correlation}
\end{table}

\begin{figure}[ht]
	\centering
	\includegraphics[width=0.8\textwidth]{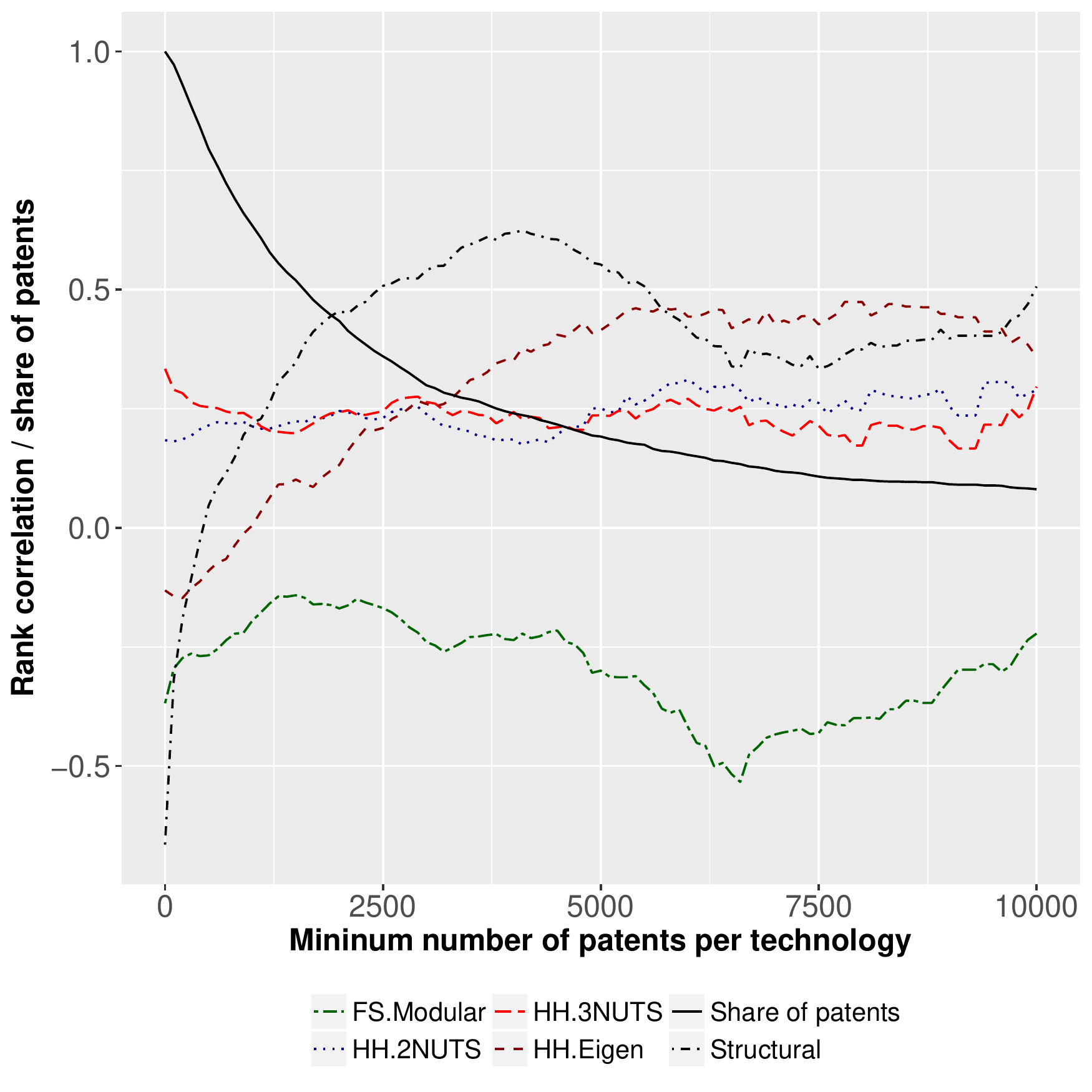}
	\caption{Correlation between technological complexity and spatial concentration (GINI coefficient)}
	\label{spatial}
\end{figure}

Figure \ref{spatial} clarifies this issue by plotting the correlation of complexity and spatial concentration for varying subsamples. More precise, I iteratively re-estimate the correlation by removing the smallest technologies from the original data whereby the technologies' minimum size (number of patents) to remain in the subsample is raised by one patent in each iteration. Accordingly, the solid lines represent the correlation coefficient given technologies of at least the according size. Additionally, the figure shows the share of patents (on all patents) still covered by the subsample (solid line). To exclude potential temporal effects, I exclusively consider the year 2010. 

The exercise has little impact on the correlation of $HH.2NUTS$ and $HH.3NUTS$ much, which remains close to 0.3. Similarly, the negative correlation of $FS.Modular$ with spatial concentration remains intact. However, the results for $HH.Eigen$ and $Structural$ change dramatically. When the smallest technologies are excluded (those with less than 350 patents in 2010) the correlation, which initially was strongly negative, becomes positive. Excluding these technologies corresponds to dropping ca. 13 \% of all patents. When excluding about 25 \% of all patents, the correlation of $Structural$ is already at the level of that of $HH.2NUTS$ and $HH.3NUTS$. It keeps increasing after this point. For $HH.Eigen$ to reach this level, almost 75 \% of all patents would have to be dropped, which suggests that spatial concentration is not a strong feature of technologies identified as complex with this measure.

In summary, the stylized fact of complex technologies concentrating in space corresponds to what can be observed when applying $HH.2NUTS$ and $HH.3NUTS$ to empirical data. However, this might be related to what is already built into this measure (see Section \ref{reflection}). The empirical results for $Structural$ also mirror this fact when excluding the smallest technologies. There is no accordance of $HH.Eigen$ and $FS.Modular$ with this stylized fact.

\subsection{Collaborative R\&D}

\begin{figure}[ht]
	\centering
	\includegraphics[width=0.8\textwidth]{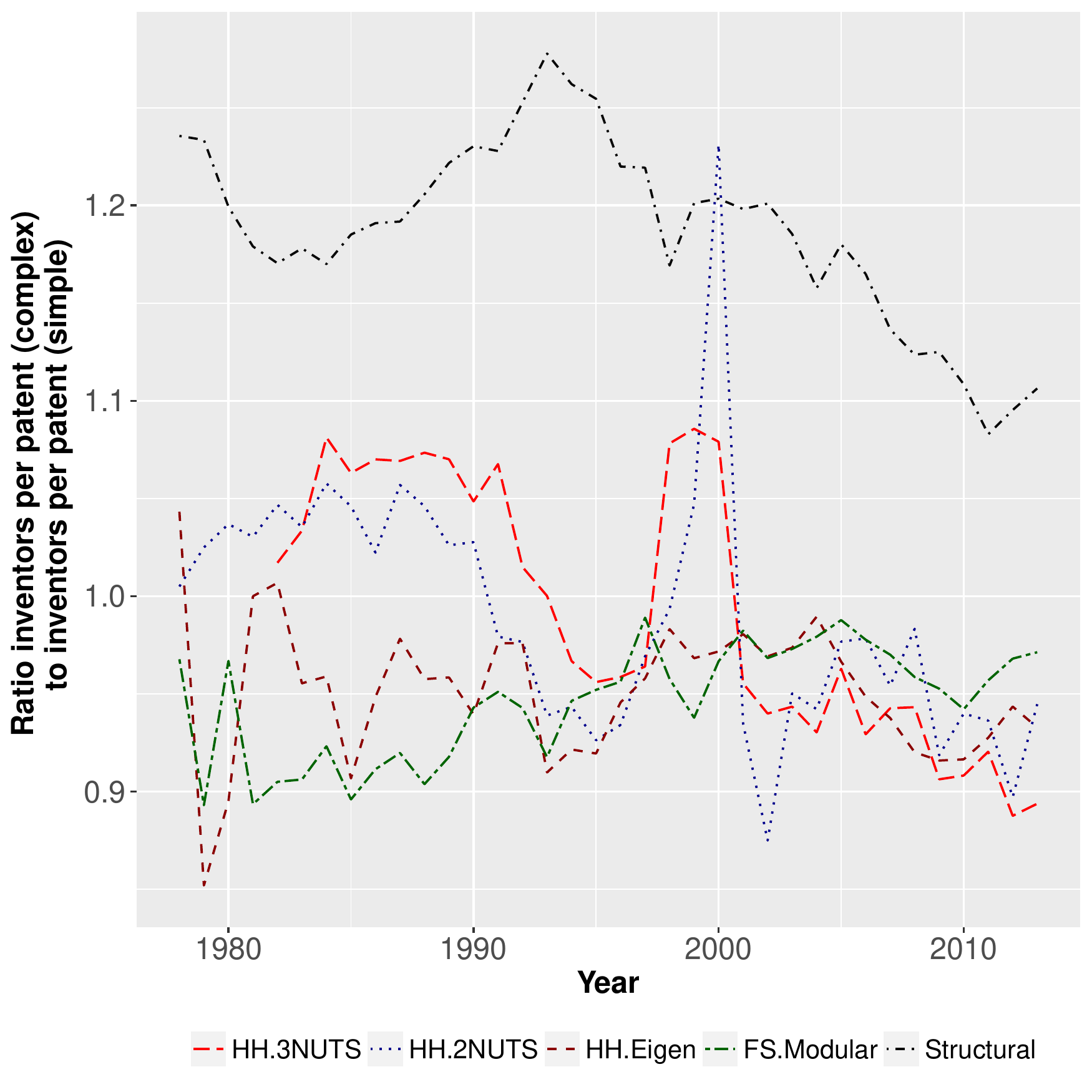}
	\caption{Correlation between complexity and inventors per patents 1980-2013}
	\label{coop}
\end{figure}

Complex technologies should show higher degrees of collaborative R\&D than simple ones. In a similar fashion as above, I explore the relation by correlating the number of inventors per patent with the five complexity measures. Figure \ref{coop} depicts this correlation over time.
The figure reveals that only $Structural$ corresponds to the stylized fact of more collaborative R\&D in complex technologies. The correlation is consistently positive and exceeds $r=0.25^{***}$ in all years. The peaking correlation between spatial concentration and $Structural$ in 1992 with a value close to 0.5 is an interesting observation that deserves more attention in future research.
All other complexity measures show negative correlations with the number of inventors per patent in extended time periods. While $HH.2NUTS$ and $HH.3NUTS$ show positive correlations until about 1993, the coefficients remains negative in most of the subsequent years. $Modular$ never manages to gain a positive correlation with spatial concentration.
Hence, it is again only $Structural$ that reflects this stylized fact.




\section{Discussion \& Conclusion} \label{conclusion}

The complexity of technologies has been measured in various ways in the past. The paper reviewed two existing empirical measures of technological complexity: the method of reflection approach by \cite{Hidalgo2009} and the difficulty of knowledge combination approach put forward by \cite{FlSo2001}. It was demonstrated that both approaches rely on critical assumptions motivating the need for alternative measures of technological complexity. Based on the work of \cite{FlSo2001} and the literature on network complexity, I proposed the new measure of $structural complexity$. It captures the complicatedness of the knowledge combinatorial process underlying technologies' advancement. 

Five distinct measures of technological complexity based on the three approaches were estimated and evaluated using European patent data for the years 1980 - 2013. I put forward four stylized facts that served as a benchmark for the evaluation: increasing (average) technological complexity over time,  complex technologies requiring more R\&D efforts, their R\&D is more collaborative, complex technologies concentrating in space, and identified complex / simple technologies meeting intuitive expectations.

Table \ref{tab:summary} summarizes the evaluation results. Only the newly introduced measure $Structural$, which captures the structural complexity of knowledge combination underlying technologies, meet all stylized facts to an acceptable degree. While it does not confirm small complex technologies being spatially concentrated, these represent a relatively small fraction of all patents.

\begin{table}[ht]
	\centering
	\resizebox{\textwidth}{!}{%
		\begin{tabular}{lccccc}
			\hline
			Stylized fact & $HH.3NUTS$  & $HH.2NUTS$ & $HH.Eigen$ & $FS.Modular$ & $Structural$ \\ 
			\hline
			Increasing complexity & No & No & Yes & Yes & Yes \\
			Larger R\&D & No & No & No & Yes & Yes \\
			Spatial concentration & Yes & Yes & No & No & Yes$^*$ \\
			Collaborative R\&D & No & No & No & No & Yes \\
			\hline
			\multicolumn{6}{l}{ $^*$ with very small technologies being exceptions}\\
			\hline
		\end{tabular}
	}
	\caption{Complexity measures and stylized facts}
	\label{tab:summary}
\end{table}

Its position is further strengthened by the empirical issues troubling the traditional measures (Table \ref{tab:issues}). When using the method of reflection approach ($HH.3NUTS$, $HH.2NUTS$, and $HH.Eigen$), the ranking of technologies in terms of complexity is found to be weakly conditional on the definition of the underlying spatial unit. Finding an appropriated spatial scale is not only a very difficult task in general, but appropriate spatial units are most likely to differ in scale between technologies. For instance, some technology's development requires spatial proximity of their underlying knowledge bases \cite[]{Balland2015} implying rather small spatial units being appropriate representations, while others do not. The latter's R\&D activities might therefore be better captured at larger spatial scales. Accordingly, any chosen scale will potentially be correct only for a share of technologies.

\begin{table}[ht]
	\centering
	\resizebox{\textwidth}{!}{%
		\begin{tabular}{lccccc}
			\hline
			Empirical issues & $HH.3NUTS$  & $HH.2NUTS$ & $HH.Eigen$ & $FS.Modular$ & $Structural$ \\ 
			\hline
			Spatial scale & Yes & Yes  &Yes & No  & No \\
			Technology size  & No & No & No & No & Yes$^*$ \\
			Computational intensity & No & No & No & No & Yes \\
			\hline
			\multicolumn{6}{l}{ $^*$ in case of small technologies}\\
			\hline
		\end{tabular}
	}
	\caption{Complexity measures and dominant empirical issues}
	\label{tab:issues}
\end{table}

Using the measure of $structural; complexity$ requires considering its strong positive correlation with technologies' size (patent counts) when technologies with few patents are considered. Moreover, by construction of the measure, the obtained complexity scores are subject to some random variation across re-estimations using the same data. These variations are however limited in scope\footnote{In non-systematic tests, I found a Pearson correlation of about $r=0.98^{***}$ across re-estimations and a rank correlations of about $r=0.91^{***}$.} and can be minimized by increasing the size of the drawn subsamples (nodes and network subsamples), though this feeds into the computational burden of the calculations. The high computational burden is another noteworthy negative feature of this measure.

Lastly, it is also worthwhile examining the technologies ranked most complex and simplest by the five measures. I therefore present the ten technologies highest ranked in terms of the five complexity measures in Tables \ref{tab:top10.HH.3NUTS}, \ref{tab:top10.HH.2NUTS}, \ref{tab:top10.HH.Eigen}, \ref{tab:top10.FS.Modular}, and \ref{tab:top10.Structural} in the Appendix. The technologies identified as being most simple are listed in Tables \ref{tab:low10.HH.2NUTS}, \ref{tab:low10.HH.3NUTS}, \ref{tab:low10.HH.Eigen}, \ref{tab:low10.FS.Modular}, and \ref{tab:low10.Structural}. Given the potentially biasing effects of small technologies, I concentrate on technologies with at least $10$ patents in the identification of the most complex ones.\footnote{The low number of patents also makes the obtained complexity scores unreliable because most of them require a sufficiently large number of empirical observations. The full rankings can be obtained from the author upon request.} It is beyond the paper's scope to discuss each and every technology in the lists but some general patterns should be mentioned. 

The lists of most complex technologies as identified by $HH.3NUTS$, $HH.2NUTS$, and $HH.Eigen$ include many technologies related to manual activities ($B23G$, $B21L$, $B27C$, $D01H$, $B25C$ $D05B$) or to natural resources ($B27B$, $B27G$). Usually, these technologies are not associated with technological complexity. According to $HH.3NUTS$ and $HH.2NUTS$ chemical technologies ($C07C$, $C07D$, $C12N$, $C07K$) are technologically simple. This is counterintuitive as chemistry is usually considered a high-tech technology involving large R\&D efforts \cite[]{scoreboard2016}. In case of $FS.Modular$, the top-ten list also includes some technologies that relate to rather simple activities ($A63C$, $A01C$,$A47J$) and hence might not considered to be complex. In contrast, the ten most simplest technologies according to this measure seem to be reasonable. It is however strongly driven by low patent numbers in these fields. The top-ten and bottom-ten lists of $Structural$ are very compelling with the size of patenting activities appearing to be a clear factor. Nevertheless, technologies ranking the in one-hundreds in terms of patent numbers, also make the top-10 list. As for $Modular$ the list of the most simple technologies is clearly driven by small patent numbers with $B61G$ ranking 484 in terms of patents among the 587 technologies with more than ten patents in 2010.

In summary, the newly proposed measure of structural complexity yields promising results and performs well with respect to the four stylized facts of technological complexity put forward in the paper.

Of course, given the lack of an objective benchmark, the presented evaluation has its limitation, which particularly relates to the four stylized facts. While the literature seems to agree on these, there is little to no supporting empirical evidence. This, of course, is in large part due to the lack of a widely-accepted complexity measure. Moreover, there might be additional stylized facts that have not been considered here. For instance, \cite{Balland2015} argue that complex technologies are likely to yield higher economic rents. This has not been included in the current assessment, as it is debatable and empirical data is missing for its assessment.

In light of this, the paper should also be seen as a call for further research and dI hope to stimulate and contribute to fruitful scientific debate on this issue.

\newpage
\section*{Appendix}

\begin{table}[ht]
	\centering
	\resizebox{\textwidth}{!}{%
	\begin{tabular}{rrrrrrrrrrrrrr}
		\hline
		& n & mean & sd & median & trimmed & mad & min & max & range & skew & kurtosis & se \\ 
		\hline
		HH.NUTS3 & 625.00 & 8.60 & 4.15 & 8.60 & 8.52 & 1.51 & 0.00 & 100.00 & 100.00 & 17.04 & 373.15 & 0.17 \\ 
		HH.NUTS2 & 625.00 & 62.48 & 11.19 & 62.84 & 62.89 & 9.22 & 0.00 & 100.00 & 100.00 & -0.61 & 2.61 & 0.45 \\ 
		HH.Eigen & 625.00 & 0.00 & 0.00 & 0.00 & 0.00 & 0.00 & 0.00 & 0.01 & 0.01 & 1.01 & 0.71 & 0.00 \\ 
		FS.Modular & 595.00 & 1.74 & 0.96 & 1.54 & 1.61 & 0.54 & 0.38 & 15.50 & 15.12 & 6.02 & 72.51 & 0.04 \\ 
		Structural & 625.00 & 7.36 & 2.54 & 7.86 & 7.63 & 2.12 & -0.00 & 13.16 & 13.16 & -1.06 & 1.18 & 0.10 \\ 
		\hline
	\end{tabular}
}
	\caption{Descriptives of complexity scores}
\label{descriptives}
\end{table}

\begin{table}[ht]
	\centering
	\begin{tabularx}{\textwidth}{llXr}
		\hline
		Rank & IPC  & Description & Patents \\ 
		\hline
540 & B23G & thread cutting working of screws, bolt heads or nuts, in conjunction therewith  &  44 \\ 
585 & B21L & making metal chains  &  11 \\ 
462 & B27C & planing, drilling, milling, turning or universal machines for wood or similar material‚  & 103 \\ 
535 & F23H & grates  &  48 \\ 
544 & B43L & articles for writing or drawing upon accessories for writing or drawing  &  39 \\ 
380 & D01H & spinning or twisting  & 202 \\ 
555 & G12B & constructional details of instruments, or comparable details of other apparatus, not otherwise provided for &  33 \\ 
424 & B25C & hand-held nailing or stapling tools manually-operated portable stapling tools  & 144 \\ 
471 & B23F & making gears or toothed racks  &  95 \\ 
419 & B27G & accessory machines or apparatus for working wood or similar materials tools for working wood or similar materials  & 154 \\ 
		\hline
\end{tabularx}
\caption{Top 10 most complex technologies with > 10 patents: HH.3NUTS}
\label{tab:top10.HH.3NUTS}
\end{table}

\begin{table}[ht]
	\centering
	\begin{tabularx}{\textwidth}{llXr}
		\hline
		Rank & IPC  & Description & Patents \\ 
		\hline
544 & B43L & articles for writing or drawing upon accessories for writing or drawing  &  39 \\ 
380 & D01H & spinning or twisting  & 202 \\ 
585 & B21L & making metal chains  &  11 \\ 
471 & B23F & making gears or toothed racks  &  95 \\ 
547 & D02J & finishing or dressing of filaments, yarns, threads, cords, ropes, or the like  &  38 \\ 
540 & B23G & thread cutting working of screws, bolt heads or nuts, in conjunction therewith  &  44 \\ 
555 & G12B & constructional details of instruments, or comparable details of other apparatus, not otherwise provided for &  33 \\ 
583 & B68B & harness devices used in connection therewith whips or the like &  12 \\ 
446 & D05B & sewing  & 121 \\ 
554 & D04C & braiding or manufacture of lace, including bobbin-net or carbonised lace braiding machines braid lace  &  33 \\ 
		\hline
	\end{tabularx}
	\caption{Top 10 most complex technologies with > 10 patents: HH.2NUTS}
	\label{tab:top10.HH.2NUTS}
\end{table}

\begin{table}[ht]
	\centering
	\begin{tabularx}{\textwidth}{llXr}
		\hline
		Rank & IPC  & Description & Patents \\ 
		\hline
251 & B25F & combination or multi-purpose tools not otherwise provided for details or components of portable power-driven tools not particularly related to the operations performed and not otherwise provided for & 470 \\ 
462 & B27C & planing, drilling, milling, turning or universal machines for wood or similar material‚  & 103 \\ 
290 & B27B & saws for wood or similar material components or accessories therefor  & 358 \\ 
279 & B25D & percussive tools & 383 \\ 
471 & B23F & making gears or toothed racks  &  95 \\ 
419 & B27G & accessory machines or apparatus for working wood or similar materials tools for working wood or similar materials  & 154 \\ 
207 & D21F & paper-making machines methods of producing paper thereon & 615 \\ 
464 & F42C & ammunition fuzes  & 103 \\ 
53 & F02M & supplying combustion engines in general with combustible mixtures or constituents thereof  & 2303 \\ 
303 & F02N & starting of combustion engines  & 335 \\ 
		\hline
	\end{tabularx}
	\caption{Top 10 most complex technologies with > 10 patents: HH.Eigen}
	\label{tab:top10.HH.Eigen}
\end{table}

\begin{table}[ht]
	\centering
	\begin{tabularx}{\textwidth}{llXr}
		\hline
		Rank & IPC  & Description & Patents \\ 
		\hline
30 & C12Q & measuring or testing processes involving enzymes or micro-organisms  & 2931 \\ 
157 & B60S & servicing, cleaning, repairing, supporting, lifting, or manoeuvring of vehicles, not otherwise provided for & 888 \\ 
2 & H04L & transmission of digital information, e.g. telegraphic communication  & 11707 \\ 
261 & A63C & skates skis roller skates design or layout of courts, rinks or the like  & 447 \\ 
112 & F24C & other domestic stoves or ranges details of domestic stoves or ranges, of general application  & 1230 \\ 
101 & B60C & vehicle tyres  & 1322 \\ 
53 & F02M & supplying combustion engines in general with combustible mixtures or constituents thereof  & 2303 \\ 
50 & A47J & kitchen equipment coffee mills spice mills apparatus for making beverages & 2320 \\ 
295 & A01C & planting sowing fertilising  & 346 \\ 
64 & A47L & domestic washing or cleaning  & 2072 \\ 
\hline
	\end{tabularx}
	\caption{Top 10 most complex technologies with > 10 patents: FS.Modular}
	\label{tab:top10.FS.Modular}
\end{table}

\begin{table}[ht]
	\centering
	\begin{tabularx}{\textwidth}{llXr}
		\hline
		Rank & IPC  & Description & Patents \\ 
		\hline
3 & A61P & specific therapeutic activity of chemical compounds or medicinal preparations & 10976 \\ 
1 & A61K & preparations for medical, dental, or toilet purposes  & 21895 \\ 
9 & H04W & wireless communication networks & 7236 \\ 
109 & B60W & conjoint control of vehicle sub-units of different type or different function control systems specially adapted for hybrid vehicles road vehicle drive control systems for purposes not related to the control of a particular sub-unit & 1275 \\ 
24 & A61Q & specific use of cosmetics or similar toilet preparations & 3476 \\ 
123 & A01P & biocidal, pest repellant, pest attractant or plant growth regulatory activity of chemical compounds or preparations & 1130 \\ 
12 & C12N & micro-organisms or enzymes medicinal preparations a61k fertilisers c05f) propagating, preserving, or maintaining micro-organisms mutation or genetic engineering culture media  & 5351 \\ 
2 & H04L & transmission of digital information, e.g. telegraphic communication  & 11707 \\ 
13 & C07K & peptides  & 4798 \\ 
6 & C07D & heterocyclic compounds  & 8045 \\
		\hline
	\end{tabularx}
	\caption{Top 10 most complex technologies with > 10 patents: Structural}
	\label{tab:top10.Structural}
\end{table}

\begin{table}[ht]
	\centering
	\begin{tabularx}{\textwidth}{llXr}
		\hline
		Rank & IPC  & Description & Nodes \\ 
		\hline 
30 & C12Q & measuring or testing processes involving enzymes or micro-organisms  & 2931 \\ 
15 & C07C & acyclic or carbocyclic compounds  & 4639 \\ 
140 & B63B & ships or other waterborne vessels equipment for shipping  & 984 \\ 
117 & A01K & animal husbandry care of birds, fishes, insects fishing rearing or breeding animals, not otherwise provided for new breeds of animals & 1186 \\ 
32 & A23L & foods, foodstuffs, or non-alcoholic beverages, not covered by subclasses  a21d  or a23b-a23j their preparation or treatment, e.g. cooking, modification of nutritive qualities, physical treatment  & 2880 \\ 
13 & C07K & peptides  & 4798 \\ 
6 & C07D & heterocyclic compounds  & 8045 \\ 
12 & C12N & micro-organisms or enzymes medicinal preparations a61k fertilisers c05f) propagating, preserving, or maintaining micro-organisms mutation or genetic engineering culture media  & 5351 \\ 
1 & A61K & preparations for medical, dental, or toilet purposes  & 21895 \\ 
3 & A61P & specific therapeutic activity of chemical compounds or medicinal preparations & 10976 \\ 
344 & F03G & spring, weight, inertia, or like motors mechanical-power-producing devices or mechanisms, not otherwise provided for or using energy sources not otherwise provided for  & 244 \\ 
		\hline
	\end{tabularx}
	\caption{10 most simple technologies with > 10 patents: HH.3NUTS}
	\label{tab:low10.HH.3NUTS}
\end{table}

\begin{table}[ht]
	\centering
	\begin{tabularx}{\textwidth}{llXr}
		\hline
		Rank & IPC  & Description & Nodes \\ 
		\hline
 619 & G06Q & data processing systems or methods, specially adapted for administrative, commercial, financial, managerial, supervisory or forecasting purposes systems or methods specially adapted for administrative, commercial, financial, managerial, supervisory or forecasting purposes, not otherwise provided for & 8813 \\ 
620 & B65D & containers for storage or transport of articles or materials, e.g. bags, barrels, bottles, boxes, cans, cartons, crates, drums, jars, tanks, hoppers, forwarding containers accessories, closures, or fittings therefor packaging elements packages & 9858 \\ 
 621 & C07K & peptides  & 11326 \\ 
 622 & F03B & machines or engines for liquids  & 754 \\ 
623 & A61P & specific therapeutic activity of chemical compounds or medicinal preparations & 27996 \\ 
624 & A61K & preparations for medical, dental, or toilet purposes  &  \\ 
625 & G07F & coin-freed or like apparatus  & 2002 \\ 
 626 & F03G & spring, weight, inertia, or like motors mechanical-power-producing devices or mechanisms, not otherwise provided for or using energy sources not otherwise provided for  & 447 \\ 
 627 & G02C & spectacles sunglasses or goggles insofar as they have the same features as spectacles contact lenses & 1025 \\ 
 628 & C12G & wine other alcoholic beverages preparation thereof  & 193 \\ 
		\hline
	\end{tabularx}
	\caption{10 most simple technologies with > 10 patents: HH.2NUTS}
	\label{tab:low10.HH.2NUTS}
\end{table}

\begin{table}[ht]
	\centering
	\begin{tabularx}{\textwidth}{llXr}
		\hline
		Rank & IPC  & Description & Nodes \\ 
		\hline
30 & C12Q & measuring or testing processes involving enzymes or micro-organisms  & 2931 \\ 
15 & C07C & acyclic or carbocyclic compounds  & 4639 \\ 
140 & B63B & ships or other waterborne vessels equipment for shipping  & 984 \\ 
117 & A01K & animal husbandry care of birds, fishes, insects fishing rearing or breeding animals, not otherwise provided for new breeds of animals & 1186 \\ 
32 & A23L & foods, foodstuffs, or non-alcoholic beverages, not covered by subclasses  a21d  or a23b-a23j their preparation or treatment, e.g. cooking, modification of nutritive qualities, physical treatment  & 2880 \\ 
13 & C07K & peptides  & 4798 \\ 
6 & C07D & heterocyclic compounds  & 8045 \\ 
12 & C12N & micro-organisms or enzymes medicinal preparations a61k fertilisers c05f) propagating, preserving, or maintaining micro-organisms mutation or genetic engineering culture media  & 5351 \\ 
1 & A61K & preparations for medical, dental, or toilet purposes  & 21895 \\ 
3 & A61P & specific therapeutic activity of chemical compounds or medicinal preparations & 10976 \\ 
344 & F03G & spring, weight, inertia, or like motors mechanical-power-producing devices or mechanisms, not otherwise provided for or using energy sources not otherwise provided for  & 244 \\ 
		\hline
	\end{tabularx}
	\caption{10 most simple technologies with > 10 patents: HH.Eigen}
	\label{tab:low10.HH.Eigen}
\end{table}

\begin{table}[ht]
	\centering
	\begin{tabularx}{\textwidth}{llXr}
		\hline
		Rank & IPC  & Description & Nodes \\ 
		\hline
551 & F15C & fluid-circuit elements predominantly used for computing or control purposes  &  36 \\ 
514 & D06C & finishing, dressing, tentering, or stretching textile fabrics  &  61 \\ 
577 & C07G & compounds of unknown constitution  &  18 \\ 
545 & H03C & modulation  &  39 \\ 
553 & B68C & saddles stirrups &  33 \\ 
562 & H05F & static electricity naturally-occurring electricity &  28 \\ 
573 & C12F & recovery of by-products of fermented solutions denaturing of, or denatured, alcohol &  22 \\ 
581 & B31C & making wound articles, e.g. wound tubes, of paper or cardboard‚ &  14 \\ 
582 & F16S & constructional elements in general structures built-up from such elements, in general &  14 \\ 
585 & B21L & making metal chains  &  11 \\ 
586 & B27J & mechanical working of cane, cork, or similar materials &  11 \\
		\hline
	\end{tabularx}
	\caption{10 most simple technologies with > 10 patents: FS.Modular}
	\label{tab:low10.FS.Modular}
\end{table}

\begin{table}[ht]
	\centering
	\begin{tabularx}{\textwidth}{llXr}
		\hline
		Rank & IPC  & Description & Nodes \\ 
		\hline
521 & E21F & safety devices, transport, filling-up, rescue, ventilation, or drainage in or of mines or tunnels &  58 \\ 
484 & B61G & couplings specially adapted for railway vehicles draught or buffing appliances specially adapted for railway vehicles &  86 \\ 
580 & B02B & preparing grain for milling refining granular fruit to commercial products by working the surface  &  14 \\ 
553 & B68C & saddles stirrups &  33 \\ 
585 & B21L & making metal chains  &  11 \\ 
586 & B27J & mechanical working of cane, cork, or similar materials &  11 \\ 
501 & G10D & stringed musical instruments wind musical instruments accordions or concertinas percussion musical instruments musical instruments not otherwise provided for  &  74 \\ 
561 & D02H & warping, beaming, or leasing &  30 \\ 
583 & B68B & harness devices used in connection therewith whips or the like &  12 \\ 
 587 & G10C & pianos, harpsichords, spinets or similar stringed musical instruments with one or more keyboards  &  11 \\  
		\hline
	\end{tabularx}
	\caption{10 most simple technologies with > 10 patents: Structural}
	\label{tab:low10.Structural}
\end{table}

\newpage
\bibliographystyle{apalike}
\bibliography{Complexity_Paper_V09.bbl}

\begin{thebibliography}{}

\bibitem[Acs et~al., 2002]{L539}
Acs, Z.~J., Anselin, L., and Varga, A. (2002).
\newblock {Patents and Innovation Counts as Measures of Regional Production of
  New Knoweldge}.
\newblock {\em Research Policy}, 31:1069--1085.

\bibitem[Albeaik et~al., 2017]{Albeaik2017}
Albeaik, S., Kaltenberg, M., Alsaleh, M., and Hidalgo, C.~A. (2017).
\newblock {Improving the Economic Complexity Index}.
\newblock {\em Arixv Working Paper}, arXiv:1707:1--21.

\bibitem[Almeida, 1996]{Almeida1996}
Almeida, P. (1996).
\newblock {Knowledge sourcing by foreign multinationals : Patent citation
  analysis in the U.S. semiconductor industry}.
\newblock {\em Strategic Management Journal}, 17(S2):155--165.

\bibitem[Arundel and Kabla, 1998]{L265}
Arundel, A. and Kabla, I. (1998).
\newblock {What percentage of innovations are patented? Empirical estimates for
  European firms}.
\newblock {\em Research Policy}, 27(2):127--141.

\bibitem[Audretsch and Feldman, 1996]{audretsch_feldman_1996}
Audretsch, D.~B. and Feldman, M. (1996).
\newblock {R{\&}D spillovers and the geography of innovation and production}.
\newblock {\em American Economic Review}, 86(4):253--273.

\bibitem[Aunger, 2010]{Aunger2010}
Aunger, R. (2010).
\newblock {Types of technology}.
\newblock {\em Technological Forecasting and Social Change}, 77(5):762--782.

\bibitem[Balland, 2016]{Balland2017}
Balland, P.~A. (2016).
\newblock {EconGeo: Computing Key Indicators of the Spatial Distribution of
  Economic Activities, R package}.
\newblock {\em https://github.com/PABalland/EconGeo}.

\bibitem[Balland and Rigby, 2017]{Balland2015}
Balland, P.-A. and Rigby, D. (2017).
\newblock {The geogrpahy and evolution of complex knowledge}.
\newblock {\em Economic Geography}, 93:1--23.

\bibitem[Bonchev and Buck, 2005]{bonchev2005}
Bonchev, D. and Buck, G. (2005).
\newblock {Quantitative measures of network complexity}.
\newblock In Bonchev, D. and Rouvrary, D., editors, {\em Complexity in
  chemistry biology and ecology}, chapter~5. Springer Verlag, New York.

\bibitem[Boschma, 2005]{L69}
Boschma, R.~A. (2005).
\newblock {Proximity and innovation: a critical assessment}.
\newblock {\em Regional Studies}, 39(1):61--74.

\bibitem[Breschi and Lenzi, 2011]{L762}
Breschi, S. and Lenzi, C. (2011).
\newblock {Net City: How co-invention networks shape inventive productivity in
  U.S. cities}.
\newblock {\em KITeS Seminarpapers}, pages 1--32.

\bibitem[Caldarelli et~al., 2012]{Caldarelli2012}
Caldarelli, G., Cristelli, M., Gabrielli, A., Pietronero, L., Scala, A., and
  Tacchella, A. (2012).
\newblock {A Network Analysis of Countries ' Export Flows : Firm Grounds for
  the Building Blocks of the Economy}.
\newblock {\em PLoS ONE}, 7(10):1--11.

\bibitem[Camagni, 1991]{L89}
Camagni, R. (1991).
\newblock {Local ``milieu'', uncertainty and innovation networks: towards a new
  dynamic theory of economic space}.
\newblock In Camagni, R., editor, {\em Innovation Networks: Spatial
  Perspectives}, pages 121--142. Belhaven Stress. London, UK and New York, USA.

\bibitem[Carbonell and Rodriguez, 2006]{Carbonell2006}
Carbonell, P. and Rodriguez, A.~I. (2006).
\newblock {Designing teams for speedy product development: The moderating
  effect of technological complexity}.
\newblock {\em Journal of Business Research}, 59(2):225--232.

\bibitem[Castaldi et~al., 2015]{Castaldi2015}
Castaldi, C., Frenken, K., and Los, B. (2015).
\newblock {Related Variety, Unrelated Variety and Technological Breakthroughs:
  An analysis of US State-Level Patenting}.
\newblock {\em Regional Studies}, 49(5):767--781.

\bibitem[Cohen and Levinthal, 1990]{L104}
Cohen, W.~M. and Levinthal, D.~A. (1990).
\newblock {Absorptive capacity: a new perspective on learning and innovation}.
\newblock {\em Administrative Science Quarterly}, 35(1):128--152.

\bibitem[Cooke, 1992]{L316}
Cooke, P. (1992).
\newblock {Regional Innovation Sytems: Competitive Regulation in the New
  Europe}.
\newblock {\em GeoForum}, 23:356--382.

\bibitem[Dalmazzo, 2002]{Dalmazzo02}
Dalmazzo, A. (2002).
\newblock {Technological complexity, wage differentials and unemployment}.
\newblock {\em Scandinavian Journal of Economics}, 104(4):515--530.

\bibitem[de~Rassenfosse and {van Pottelsberghe de la Potterie},
  2009]{DeRassenfosse2009}
de~Rassenfosse, G. and {van Pottelsberghe de la Potterie}, B. (2009).
\newblock {A policy insight into the R{\&}D-patent relationship}.
\newblock {\em Research Policy}, 38(5):779--792.

\bibitem[Dehmer et~al., 2009]{Dehmer2009}
Dehmer, M., Barbarini, N., Varmuza, K., and Graber, A. (2009).
\newblock {A large scale analysis of information-theoretic network complexity
  measures using chemical structures}.
\newblock {\em PLoS ONE}, 4(12):20--26.

\bibitem[Dehmer and Mowshowitz, 2011]{Dehmer2011}
Dehmer, M. and Mowshowitz, A. (2011).
\newblock {A history of graph entropy measures}.
\newblock {\em Information Sciences}, 181(1):57--78.

\bibitem[Emmert-Streib and Dehmer, 2012]{Emmert-Streib2012}
Emmert-Streib, F. and Dehmer, M. (2012).
\newblock {Exploring statistical and population aspects of network complexity}.
\newblock {\em PLoS ONE}, 7(5).

\bibitem[Fai and {Von Tunzelmann}, 2001]{Fai2001}
Fai, F. and {Von Tunzelmann}, N. (2001).
\newblock {Industry-specific competencies and converging technological systems:
  Evidence from patents}.
\newblock {\em Structural Change and Economic Dynamics}, 12(2):141--170.

\bibitem[Feldman, 1994]{L19}
Feldman, M. (1994).
\newblock {\em {The Geography of Innovation}}.
\newblock Economics of Science, Technology and Innovation, Vol. 2, Kluwer
  Academic Publishers, Dordrecht.

\bibitem[{Fernandez Donoso}, 2017]{FernandezDonoso2017}
{Fernandez Donoso}, J. (2017).
\newblock {A simple index of innovation with complexity}.
\newblock {\em Journal of Informetrics}, 11(1):1--17.

\bibitem[Fleming and Frenken, 2007]{FlFr2007}
Fleming, L. and Frenken, K. (2007).
\newblock {The evolution of inventor networks in the Silicon Valley and Boston
  regions}.
\newblock {\em Advances in Complex Systems}, 10(1):53--71.

\bibitem[Fleming and Sorenson, 2001]{FlSo2001}
Fleming, L. and Sorenson, O. (2001).
\newblock {Technology as a complex adaptive system: evidence from patent data}.
\newblock {\em Research Policy}, 30(7):1019--1039.

\bibitem[Fleming and Sorenson, 2004]{Fleming2004}
Fleming, L. and Sorenson, O. (2004).
\newblock {Science as a map in technological search}.
\newblock {\em Strategic Management Journal}, 25(8-9):909--928.

\bibitem[Florida, 1995]{L87}
Florida, R. (1995).
\newblock {Toward the Learning Region}.
\newblock {\em Futures}, 27:527--536.

\bibitem[Frenken et~al., 2005]{Frenkenetal2005}
Frenken, K., H{\"{o}}lzl, W., and de~Vor, F. (2005).
\newblock {The citation impact of research collaborations: the case of European
  biotechnology {\&} applied microbiology (1988–2002)}.
\newblock {\em Journal of Engineering and Technology Management},
  22(1-2):9--30.

\bibitem[Frenken et~al., 2007]{L702}
Frenken, K., van Oort, F.~G., and Verburg, T. (2007).
\newblock {Related variety, unrelated variety and regional economic growth}.
\newblock {\em Regional Studies}, 41(5):685--697.

\bibitem[Griffin, 1997]{Griffin1997}
Griffin, A. (1997).
\newblock {The Effect of Project and Process Characteristics on Product
  Development Cycle Time}.
\newblock {\em Journal of Marketing Researcharketing Research}, 34(1):24--35.

\bibitem[Griliches, 1990]{L589}
Griliches, Z. (1990).
\newblock {Patent statistics as economic indicators: A survey}.
\newblock {\em Journal of Economic Literature}, 28:1661--1701.

\bibitem[Hagedoorn et~al., 2000]{Hage00}
Hagedoorn, J., Link, A.~N., and Vonortas, N.~S. (2000).
\newblock {Research partnerships}.
\newblock {\em Research Policy}, 29(4-5):567--586.

\bibitem[H{\"{a}}gerstrand, 1967]{hagerstrand1967}
H{\"{a}}gerstrand, T. (1967).
\newblock {\em {Innovation diffusion as a spatial process}}.
\newblock University of Chicago Press, Chicago.

\bibitem[Harhoff et~al., 1999]{HNSV1999}
Harhoff, D., Narin, F., Scherer, F.~M., and Vopel, K. (1999).
\newblock {Citation Frequency and the Value of Patented Inventions}.
\newblock {\em Review of Economics and Statistics}, 81(3):511--515.

\bibitem[Hidalgo and Hausmann, 2009]{Hidalgo2009}
Hidalgo, A. and Hausmann, R. (2009).
\newblock {The building blocks of economic complexity}.
\newblock {\em PNAS}, 106(26):10570--10575.

\bibitem[Hidalgo, 2015]{Hidalgo2015}
Hidalgo, C.~A. (2015).
\newblock {\em {Why Information Grows: The Evolution of Order, from Atoms to
  Economies}}.
\newblock Basic Books, New York.

\bibitem[Hoekman et~al., 2009]{Hoekman_et_al_2009}
Hoekman, J., Frenken, K., and Oort, F. (2009).
\newblock {The geography of collaborative knowledge production in Europe}.
\newblock {\em The Annals of Regional Science}, 43(3):721--738.

\bibitem[Howitt, 1999]{Howitt1999}
Howitt, P. (1999).
\newblock {Steady Endogenous Growth with Population and R . {\&} D . Inputs
  Growing}.
\newblock {\em Journal of Political Economy}, 107(4):715--730.

\bibitem[IRI, 2016]{scoreboard2016}
IRI (2016).
\newblock {\em {The 2016 EU Industrial R{\&}D Investment Scoreboard}}.
\newblock Publications Office of the European Union, Luxembourg.

\bibitem[Jaffe, 1989a]{Jaff1989a}
Jaffe, A.~B. (1989a).
\newblock {Characterizing the "technological position" of firms, with
  application to quantifying technological opportunity and research
  spillovers}.
\newblock {\em Research Policy}, 18(2):87--97.

\bibitem[Jaffe, 1989b]{L184}
Jaffe, A.~B. (1989b).
\newblock {Real effects of academic research}.
\newblock {\em American Economic Review}, 79(5):957--970.

\bibitem[Katz and Martin, 1997]{KatzMartin1997}
Katz, J.~S. and Martin, B.~R. (1997).
\newblock {What is research collaboration?}
\newblock {\em Research Policy}, 26(1):1--18.

\bibitem[Kauffmann, 1993]{Kauffmann1993}
Kauffmann, S. (1993).
\newblock {\em {The origins of order: Self-organization and selection in
  evolution.}}
\newblock Oxford University Press., New York.

\bibitem[Kogler et~al., 2013]{Kogler2013}
Kogler, D.~F., Rigby, D.~L., and Tucker, I. (2013).
\newblock {Mapping Knowledge Space and Technological Relatedness in US Cities}.
\newblock {\em European Planning Studies}, 21(9):1374--1391.

\bibitem[Markusen, 1996]{L527}
Markusen, A. (1996).
\newblock {Sticky Places in slippery Space: A Typology of Industrial
  Districts}.
\newblock {\em Economic Geography}, 72(3):293--313.

\bibitem[Meyer and Bhattacharya, 2004]{Meyer2004}
Meyer, M. and Bhattacharya, S. (2004).
\newblock {Commonalities and differences between scholarly and technical
  collaboration An exploration of co-invention and co-authorship analyses}.
\newblock {\em Scientometrics}, 61(3):443--456.

\bibitem[Mueller et~al., 2011]{Mueller2011}
Mueller, L. A.~J., Kugler, K.~G., Graber, A., Emmert-Streib, F., and Dehmer, M.
  (2011).
\newblock {Structural measures for network biology using QuACN.}
\newblock {\em BMC bioinformatics}, 12:492.

\bibitem[Nelson, 1982]{L164}
Nelson, R.~R. (1982).
\newblock {The Role of Knowledge in R{\&}D Efficiency}.
\newblock {\em The Quarterly Journal of Economics}, 97(3):453--470.

\bibitem[Nelson and Winter, 1982]{NelsWint82}
Nelson, R.~R. and Winter, S.~G. (1982).
\newblock {The Schumpeterian tradeoff revisited}.
\newblock {\em American Economic Review}, 72(1):114--132.

\bibitem[Nooteboom, 2000]{Noot2000}
Nooteboom, B. (2000).
\newblock {Learning by interaction: absorptive capacity, cognitive distance and
  governance}.
\newblock {\em Journal of Management and Governance}, 4:69--92.

\bibitem[Pintea and Thompson, 2007]{Pintea2007}
Pintea, M. and Thompson, P. (2007).
\newblock {Technological complexity and economic growth}.
\newblock {\em Review of Economic Dynamics}, 10(2):276--293.

\bibitem[Prencipe, 2000]{Prencipe2000}
Prencipe, A. (2000).
\newblock {Breadth and depth of technological capabilities in CoPS: the case of
  the aircraft engine control system}.
\newblock {\em Research Policy}, 29:895--911.

\bibitem[Rogers, 1995]{Roge95}
Rogers, E.~M. (1995).
\newblock {\em {The Diffusion of Innovation, 4th ed.}}
\newblock Free Press, New York.

\bibitem[Romer, 1990]{L153}
Romer, P.~M. (1990).
\newblock {Endogenous technological change}.
\newblock {\em Journal of Political Economy}, 98:71--102.

\bibitem[Saxenian, 1994]{Saxe94}
Saxenian, A. (1994).
\newblock {\em {Regional Advantage - cluture and Competition in Silicon Valley
  and Route 128}}.
\newblock Harvard University Press, Cambridge.

\bibitem[Schmoch et~al., 2003]{L392}
Schmoch, U., Laville, F., Patel, P., and Frietsch, R. (2003).
\newblock {Linking technology areas to industrial sectors}.
\newblock {\em Final Report to the European Commission, DG Research, Karlsruhe,
  Paris, Brighton}.

\bibitem[Shannon, 1948]{Shannon1948}
Shannon, C.~E. (1948).
\newblock {A mathematical theory of communication}.
\newblock {\em The Bell System Technical Journal}, 27(July 1928):379--423.

\bibitem[Simon, 1962]{Simon1962}
Simon, H.~A. (1962).
\newblock {The architecture of complexity}.
\newblock {\em Proceedings of the american philosphical society},
  106(6):467--482.

\bibitem[Singh, 1997]{singh1997}
Singh, K. (1997).
\newblock {The impact of technological complexity and interfirm cooperation on
  business survival}.
\newblock {\em Academy of Management Journal}, 40(2):339--367.

\bibitem[Sorenson, 2005]{Sorenson2005}
Sorenson, O. (2005).
\newblock {Social Networks, Informational Complexity and Industrial Geography}.
\newblock In Fornahl, D., Zellner, C., and Audretsch, D., editors, {\em The
  Role of Labour Mobility and Informal Networks for Knowledge Transfer},
  chapter~5, pages 79--96. Springer Science+Business Media Inc., Boston.

\bibitem[Sorenson et~al., 2006]{SoRF2006a}
Sorenson, O., Rivkin, J.~W., and Fleming, L. (2006).
\newblock {Complexity, networks and knowledge flow}.
\newblock {\em Research Policy}, 35(7):994--1017.

\bibitem[Tacchella et~al., 2012]{Tacchella2012}
Tacchella, A., Cristelli, M., Caldarelli, G., Gabrielli, A., and Pietronero, L.
  (2012).
\newblock {A New Metrics for Countries' Fitness and Products' Complexity}.
\newblock {\em Scientific Reports}, 2:1--4.

\bibitem[Teece, 1992]{Teec92}
Teece, D. (1992).
\newblock {Competition, cooperation, and innovation: Organizational
  arrangements for regimes of rapid technological progress}.
\newblock {\em Journal of Economic Behavior and Organization}, 18(1):1--25.

\bibitem[Teece, 1977]{teece1977}
Teece, D.~J. (1977).
\newblock {Technology Transfer by Multinational Firms : The Resource Cost of
  Transferring Technological Know-How}.
\newblock {\em Economic Journal}, 87:242--261.

\bibitem[{Ter Wal}, 2013]{TerWal2013}
{Ter Wal}, A. L.~J. (2013).
\newblock {The dynamics of the inventor network in german biotechnology:
  Geographic proximity versus triadic closure}.
\newblock {\em Journal of Economic Geography}, 14(3):589--620.

\bibitem[Trajtenberg, 1990]{Traj90}
Trajtenberg, M. (1990).
\newblock {A penny for your quotes: Patent citations and the value of
  innovations}.
\newblock {\em RAND Journal of Economics}, 20:172--187.

\bibitem[von Hippel, 1994]{L135}
von Hippel, E. (1994).
\newblock {``Sticky Information'' and the Locus of Problem Solving:
  Implications for Innovation}.
\newblock {\em Management Science}, 40(4):429--439.

\bibitem[Wagner-Doebler, 2001]{WagnerDoebler2001}
Wagner-Doebler, R. (2001).
\newblock {Continuity and discontinuity of collaboration behaviour since 1800 -
  from a bibliometric point of view}.
\newblock {\em Scientometrics}, 52(3):503--517.

\bibitem[Wiener, 1947]{Wiener1947}
Wiener, H. (1947).
\newblock {Structural determination of paraffin boilin points}.
\newblock {\em Journal of the American Chemical Society}, 69(1):17--20.

\bibitem[Wikipedia, 2017]{wiki_lines_code_2017}
Wikipedia (2017).
\newblock {Source lines of code}.

\bibitem[{Woo Kim}, 2015]{WooKim2015}
{Woo Kim}, B. (2015).
\newblock {Economic Growth: Education vs. Research}.
\newblock {\em Journal of Global Economics}, 03(04).

\bibitem[Wuchty et~al., 2007]{WuJU2007}
Wuchty, S., Jones, B.~F., and Uzzi, B. (2007).
\newblock {The Increasing Dominance of Teams in Production of Knowledge}.
\newblock {\em Science}, 316(5827):1036--1039.

\bibitem[Zander and Kogut, 1995]{L253}
Zander, U. and Kogut, B. (1995).
\newblock {Knowledge and the Speed of the Transfer and Imitation of
  Organizational Capabilities: An Empirical Test}.
\newblock {\em Organization Science}, 6(1):76--91.

\end{thebibliography}
\end{document}